\documentclass[preprint]{aastex}

\received{}
\accepted{}

\slugcomment{\apj, accepted}

\lefthead{Richards et al.}
\righthead{QSOALSs as a Function of Orientation Measures}

\begin{document}

\title{Quasar Absorption Lines as a Function of Quasar Orientation Measures} 

\author{Gordon T. Richards\altaffilmark{1}}
\affil{Department of Astronomy and Astrophysics, University of Chicago, 5640 S. Ellis Avenue, Chicago, IL 60637}
\email{richards@oddjob.uchicago.edu}

\author{S. A. Laurent-Muehleisen and Robert H. Becker}
\affil{UC-Davis and IGPP/LLNL, L-413, 7000 East Ave., Livermore, CA 94550}
\email{slauren,bob@igpp.ucllnl.org}

\and

\author{Donald G. York\altaffilmark{2}}
\affil{Department of Astronomy and Astrophysics, University of Chicago, 5640 S. Ellis Avenue, Chicago, IL 60637}
\email{don@oddjob.uchicago.edu}

\altaffiltext{1}{Current address: Department of Astronomy and
Astrophysics, The Pennsylvania State University, University Park, PA
16802}
\altaffiltext{2}{Also, Enrico Fermi Institute, 5640. S. Ellis Avenue, Chicago, IL 60637} 
\begin{abstract}

We present high resolution radio data at 3.5 and 20\,cm from the VLA
in the A configuration for 144 quasars with known \ion{C}{4}
absorption line properties.  Using these measurements, we compare and
contrast a number of quasar orientation indicators.  These quantities
are used to study the velocity distribution of \ion{C}{4} absorption
lines as a function of quasar orientation measures.  That there is an
excess of narrow, high-velocity \ion{C}{4} absorbers in flat-spectrum
quasars as compared to steep-spectrum quasars is confirmed.
Orientation indicators that are not based upon spectral indices
(e.g. $R_V$ and the core-to-lobe ratio) do not show the same effect.
These seemingly contradictory results may be reconciled if the
absorber distribution is not strictly a function of orientation, but
rather a function of intrinsic radio properties that may or may not be
good orientation indicators.
 
\end{abstract}


\section{Introduction}

There is a wealth of classes of extragalactic radio sources. The idea
that some or all of these sources come from the same parent population
(i.e. the unified model) has been around for quite some time
\citep{rr77,sr79}.  Specifically, it is thought that the
classification of an object may have as much to do with our point of
view as with the physics of the source.  If the parent population is
intrinsically non-spherically symmetric, then it is clear how such
orientation effects might come about: sources drawn from the same
parent population will look different depending on the observer's
vantage point.

For example, it has been suggested that type 1 and type 2 Seyfert
galaxies differ only by orientation.  The intrinsically non-spherical
element in this model is a thick molecular torus that shields our view
of Seyfert 2 broad line regions.  This model gained widespread
acceptance when \citet{am85} discovered hidden broad lines in NGC
1068's polarized (scattered) light spectrum.  Various models using
beamed radio jet emission have been proposed to unify flat-spectrum
radio quasars and steep-spectrum radio galaxies, but the jury is still
out on the exact form this model must take \citep[e.g.,][]{ob82,pu92}.
For the purposes of this paper, it suffices that the average
flat-spectrum quasar is probably viewed closer to the jet axis than
the average steep-spectrum quasar, though such claims cannot be made
for comparisons of individual objects.

The purpose of this paper is twofold.  In the first part of the paper,
we present new 3.5 and 20\,cm VLA data (in the A configuration) taken
for 120 of the sources from \citet[hereafter Paper I]{ryy+99} and 24
sources from \citet[hereafter Paper II]{ric00}.  Using these new radio
maps, we derive the core and the total spectral indices between 3.5
and 20\,cm ($\alpha_{3.5}^{20}$), the core-to-lobe ratio
(C)\footnote{The core-to-lobe ratio is denoted as $C$ rather than $R$
to avoid confusion with the use of $R$ as the radio-loudness
parameter}, and the ratio of the radio luminosity (at 20\,cm) to the
optical luminosity ($R_V$).  Each of these parameters has been used as
an indicator of the line-of-sight orientation of quasars.  We discuss
the utility of these parameters as such, with emphasis on testing the
hypothesis that radio spectral indices correlate with the
line-of-sight orientation of quasars.

The second part of the paper addresses the results of Paper I, where
we found that the distribution of \ion{C}{4} absorption lines along
the line of sight to QSOs may be a function of various QSO properties.
This absorption line analysis in Paper I suffered from a number of
inadequacies resulting from inhomogeneous radio data.  The 6 and
20\,cm radio data all were taken at different epochs and may be
affected by variability.  In addition, these data were taken from
catalogs of data from both the Green Bank 6 and $20\,{\rm cm}$ surveys
\cite[respectively]{bwe91,wb92} and were mixed with $20\,{\rm cm}$
data from the NRAO VLA Sky Survey (NVSS; \citealt{ccg+98}).
Furthermore, the flux density limit of the Green Bank surveys is such
that only relatively bright radio sources have measured spectral
indices.  As a result, we felt it was necessary to reobserve some of
the objects from Paper I at lower flux density limits, at higher
spatial resolution, with a longer baseline, and simultaneously in two
bands.

Utilizing this new radio data, we can confirm an apparent excess of
narrow, apparently high-velocity \ion{C}{4} absorbers seen in
flat-spectrum quasars as compared to steep-spectrum quasars.  This
result is surprising in that the vast majority of these absorption
lines are thought to be caused by galaxies between the observer and
the QSO: intrinsic QSO properties should not affect their
distribution.  Though we presume that the radio properties discussed
herein are good indicators of quasar orientations, the results
presented here do not depend on this assumption.  It suffices that
these radio properties are indicators of intrinsic differences in the
quasars.

This paper is divided as follows.  \S 2 describes the targets, the
radio data and the reduction of said radio data.  Quasar orientation
measures are discussed in \S 3.  \S 4 gives a discussion of both
high-velocity and low-velocity \ion{C}{4} absorption lines systems as
a function of the properties discussed in \S 3. Finally, \S 5 presents
our conclusions.

\section{The Data}

The objects chosen for observations with the VLA in the A
configuration were selected from the revised catalog of absorption
line QSOs \citep{yyc+91,v+00} as follows.  First, we required that
each object be detected by the low resolution Green Bank or NVSS
Surveys, which had $20\,{\rm cm}$ flux density limits of $\sim19\,{\rm
mJy}$ and $\sim2.5\,{\rm mJy}$, respectively.  Since the primary goal
of this work is to study the distribution of \ion{C}{4} absorbers in
radio-detected quasars, we only observed objects with redshifts larger
than $1.5$, so that \ion{C}{4} emission is observable in ground-based
spectra.  In addition, sources with declination less than -27.5
degrees declination were rejected.  Though the VLA can observe further
south than this limit, we chose not to observe such quasars so that
followup spectroscopy could be conducted from the Northern Hemisphere.

Data were taken on 31 July and 9 August 1999, with the
VLA\footnote{The VLA and NRAO are operated by Associated Universities,
Inc., under cooperative agreement with the National Science
Foundation.} in the A-configuration.  Snapshot observations (45 sec --
7.5 min, depending on expected source brightness) were performed
sequentially on each target at both 1.4 and 8.4\,GHz (20 and 3.5\,cm).
Absolute flux calibration was established with short observations of
3C286 and 3C147, respectively, on the two dates of observation.  Phase
calibrators were also observed periodically throughout both runs.

Data were reduced in the standard iterative manner using the 15APR98
version of the AIPS software package.  The data were initially mapped,
cleaned and self-calibrated using the task MAPIT; as warranted by
these initial maps, some sources were further processed using the
tasks IMAGR and CALIB.  Clipping of especially bad data points was
done as necessary on a source-by-source basis.  Final source positions
and peak fluxes were determined using the task JMFIT, whereas total flux
densities were determined using the task TVSTAT to encircle all
components associated with each radio source.

In Paper I there were 296 QSOs with measured radio spectral indices.
Of these, 174 meet the redshift and declination limit for our present
study.  We observed 144 quasars during the course of this work, of
which 120 were drawn from the sample of 174.  The remaining 24 objects
were drawn from Paper II.  For 99 of these quasars, we obtained good
3.5 and 20\,cm detections and were able to match the position of the
radio core to the optical position such that we were able to determine
the core spectral index.  An additional 20 objects had good 20\,cm
detections (that coincided with the optical position) but no 3.5\,cm
detections.  For those cases where we could not clearly identify the
radio core, we still determined the total flux densities and total
spectral indices.

The flux density limits of the original Green Bank data are such that
only relatively bright radio sources have accurately measured radio
spectral indices.  The lack of fainter radio sources in our sample may
constitute a bias in the data from Paper I.  As such, we were careful
in our new observations to include objects that are as faint as the
flux density limit of the NVSS.  The inclusion of fainter radio
sources thus enables us to increase the number of quasars with
accurate spectral indices, while helping to remove any bias incurred
by the study of only the brightest sources.  The total number of faint
sources is still a significant minority; however, it is hoped that the
inclusion of fainter objects will reduce any luminosity bias in the
data.

These new data represent an additional significant improvement on that
of Paper I in that we now have simultaneous data in two bands
(eliminating the effects of source variability).  In addition, these
radio maps are of higher spatial resolution than the previous radio
data.  Since core spectral indices may be correlated with quasar
orientation better than total (core+extended) spectral indices, it is
hoped that these new radio data will solidify our absorption line
analysis.

In Table~\ref{tab:tab1} we present the observed properties of the
quasars studied in this work.  Given are 1) the name, 2) the right
ascension in J2000 coordinates, 3) the declination, 4) the emission
redshift, 5) an optical magnitude (often, but not always $V$), 6) the
6\,cm Green Bank flux density (mJy) at low resolution, 7) the 20\,cm
Green Bank flux density (mJy) at low resolution.  Also given are flux
densities from the VLA in the A configuration including: 8) the peak
3.5\,cm flux density, 9) the total 3.5\,cm flux density, 10) the peak
20\,cm flux density, and 11) the total 20\,cm flux density.  In
addition to these quantities, we also determined an RMS error for each
map and the uncertainty in the peak flux density measurements.  Total
errors on the peak flux density were determined by taking the sum in
quadrature of the RMS noise in the map, the error in the fit as
determined by JMFIT and 5\% of the peak flux density.  For the
majority of the sources, this error will be dominated by last of these
terms.  The resolution of the VLA A configuration data is
approximately $0.7\arcsec$ for the 20\,cm (L band) measurements and
$0.1\arcsec$ for the 3.5\,cm (X band) measurements.

Table~\ref{tab:tab2} gives parameters derived from those in
Table~\ref{tab:tab1}.  The columns are as follows: 1) the name, 2) the
low resolution spectral index between 6 and 20\,cm
($\alpha_6^{20}$)\footnote{Spectral indices are defined throughout
according to $f_{\nu} \propto \nu^{\alpha}$, such that steep-spectrum
sources have more negative spectral indices.} as determined from
columns 6 and 7 in Table~\ref{tab:tab1}, 3) the core spectral index
between 3.5 and 20\,cm ($\alpha_{3.5}^{20}$) using the values in
columns 8 and 10 in Table~\ref{tab:tab1}, 4) the error in
$\alpha_{3.5}^{20}$ (core), 5) the total ($\alpha_{3.5}^{20}$)
spectral index as found from columns 9 and 11 in Table~\ref{tab:tab1},
6) the ratio of the core radio luminosity to the optical luminosity
($R_V$), which is given by
\begin{equation}
R_V = \log(f_{20})-(1.0+\alpha)\log(1.0+z)-26.0+0.4(V+48.6),
\end{equation}
where we have assumed an optical spectral index of $-1$ and have not
converted the 20\,cm flux density to the more commonly used 6\,cm flux
density, 7) the core-to-lobe ratio at 20\,cm (which we will label as
``$C_{20}$''), given by 
\begin{equation}
C_{20} = \left(f_{20hi}\left(1+z_{em}\right)^{\left(\alpha_{core}-1\right)}\right)/\left(f_{20lo}-f_{20 hi}\right),
\end{equation}
 where we have taken the spectral index of the extended emission to be
$-1$, 8) the core 20\,cm luminosity (${\rm ergs\,s^{-1}\,Hz^{-1}}$),
9) the total 20\,cm luminosity (${\rm ergs\,s^{-1}\,Hz^{-1}}$), and
10) the absolute optical magnitude.  For those quantities that are
dependent upon cosmology, we have used $q_{\rm o} = 0.5$ and $H_{\rm
o} = 65\,{\rm km\,s^{-1}\,Mpc^{-1}}$.

\placetable{tab:tab1}
\placetable{tab:tab2}

\section{Quasar Orientation Measures}

A number of parameters are currently being used as orientation
indicators for radio-detected quasars and galaxies.  The accuracy of a
given parameter is typically evaluated by comparing to another such
parameter --- a process that is unappealing in its circularity.
Barring more direct data that could only come from in situ inspection,
it may not be possible to do much better.  Fortunately, the purpose
for which we need these orientation measures (comparing the absorption
line velocity distribution between quasars with different intrinsic
properties) does not depend on their being good orientation measures
per se, but rather that they are correlated with apparent intrinsic
differences projected along our line of sight.

Orientation measures include radio spectral indices (both core and
total), core-to-lobe ratios, core-to-optical luminosity ratios, bend
angles, core-hot spot distance ratios, etc.  In this analysis, we will
compare and contrast the use of core and total radio spectral indices
along with $R_V$ and $C_{20}$ (the core-to-lobe ratio at 20\,cm) as
orientation indicators.  The basis set for this study is the 144
$z>1.5$ quasars with known absorption line properties for which we now
have 3.5 and 20\,cm maps from the VLA in the A configuration.

\begin{figure}[p]
\epsscale{1.0}
\plotone{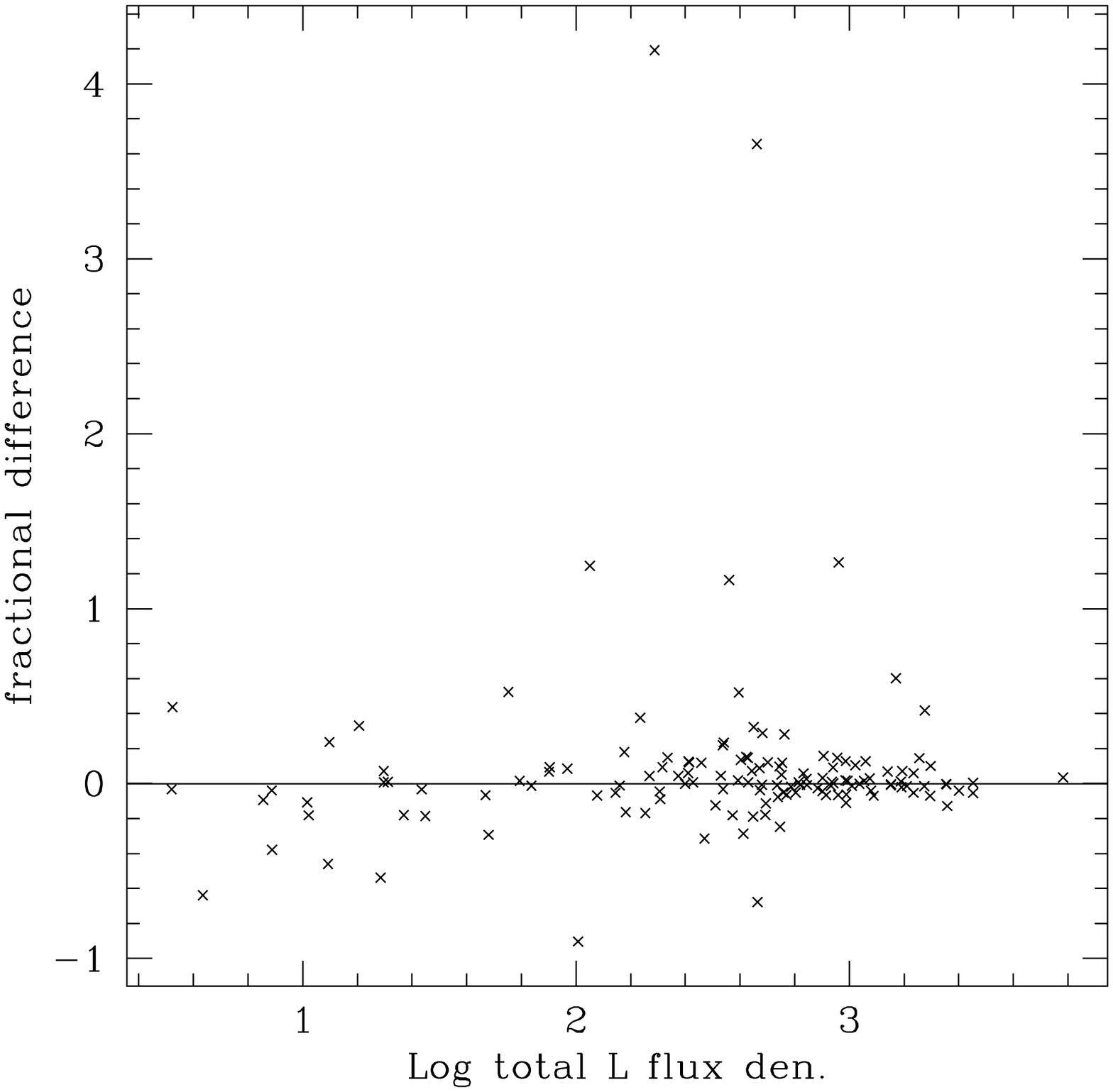}
\caption{Fractional difference between low resolution 20cm flux density and 
high resolution 20cm flux density versus the log of the high
resolution flux density. (Large positive values indicate excess flux
at lower resolution.) \label{fig:fig1}}
\end{figure}

One of the primary concerns that we had with our analysis in Paper I
was the use of radio data taken at different epochs, meaning our
analysis could have been influenced by variability in our data.
Without simultaneous measurements, we were forced to assume that our
data, on average, were unaffected by variability.  With our new
simultaneous 3.5 and 20\,cm maps, we can now address this problem.
Figure~\ref{fig:fig1} shows the fractional difference between the
total 20\,cm flux density as measured by Green Bank or NVSS (the
measurement used in Paper I) and our high resolution VLA maps.  This
difference is plotted against the log of the total 20\,cm flux density
from our new measurements.  Large differences are indicative of either
variability or resolution effects; large positive values on the y-axis
indicate excess flux in the lower resolution data. The fact that most
of our sources have differences less than about 20\% means that
variability should not strongly influence our results.  However, it is
clear that some sources have varied by 50\% or more.  The fact that
the measurements were taken at different resolutions may contribute to
these differences.

\placefigure{fig:fig1}

\begin{figure}[p]
\epsscale{1.0}
\plotone{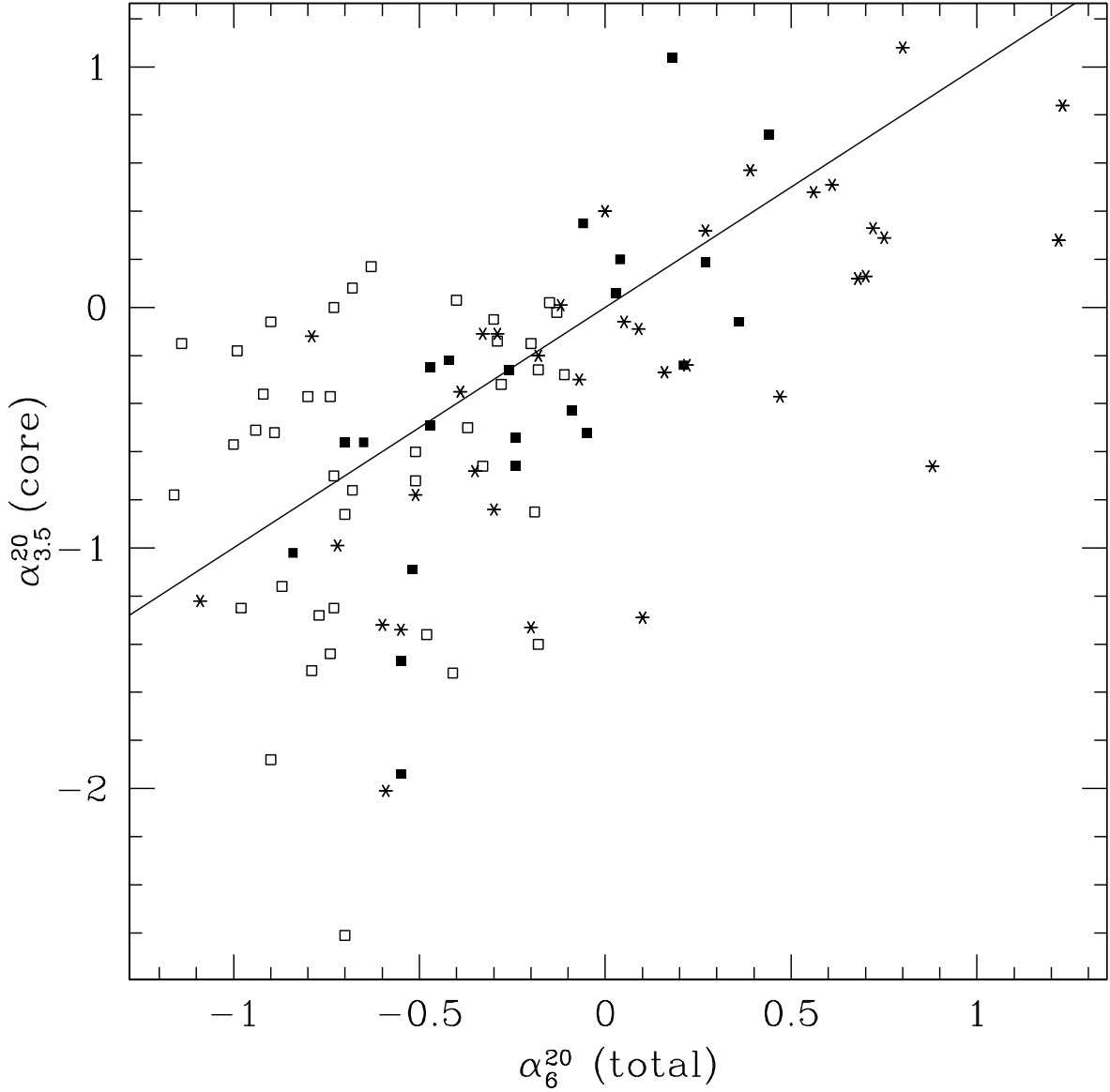}
\caption{Core spectral index ($f_{\nu} \propto \nu^{\alpha}$) between
the 3.5 and 20\,cm from our A configuration data versus the total
spectral index between 6 and 20\,cm from lower resolution data.
Closed squares are unresolved point sources (core-to-lobe ratio
greater than unity).  Open squares are lobe-dominated sources
(core-to-lobe ratio less than unity), and stars are objects with
unreliable core-to-lobe ratios.  The solid line traces the values
where the spectral indices are equal: it is not a fit to the data.
Typical errors in $\alpha_{3.5}^{20}$ (core) are
$\pm0.1$. \label{fig:fig2}}
\end{figure}

Since the A configuration data were taken simultaneously, we assume
that the total spectral indices calculated from the VLA data are
``correct''.  Variability will then either conspire to yield incorrect
spectral indices in the low resolution data (taken at different
epochs), or will have no apparent effect if the low resolution
observations happen to give the ``correct'' (or nearly correct)
spectral index.  Since flat-spectrum sources are more likely to be
variable than steep-spectrum sources, we would expect that objects
with flat core spectra (measured simultaneously at two frequencies)
would exhibit indices that depart significantly from $\alpha \approx
0$ when computed from non-simultaneous radio data.  However, from
Figure~\ref{fig:fig2} it appears that this effect is not dramatic.
Only 18 of the 45 objects with flat ($-0.4 < \alpha <
0.4$)\footnote{The typical dividing line between steep- and
flat-spectrum sources is $\alpha = -0.5$.  Here we use $\alpha =
-0.4$, since that is the median of our core spectral index sample.}
simultaneously-measured core spectral indices, have non-simultaneous
(total flux) spectral indices that fall outside this range.

\placefigure{fig:fig2}

In Figure~\ref{fig:fig2} (and similar graphs), open squares are those
objects with 20\,cm core-to-lobe ratios ($C_{20}$) less than unity.
Filled squares are objects with core-to-lobe ratios greater than
unity, whereas stars are those objects for which the core-to-lobe
ratio is suspect or otherwise not available.  Spectral indices were
calculated according to $f_{\nu} \propto \nu^{\alpha}$ such that
steep-spectrum sources have more negative spectral indices.  Though it
is possible that variability is the cause of the ``outliers''
described in the paragraph above, it must also be kept in mind that
some of the ``variability'' observed in the data is likely a result of
the different resolutions used to measure the core and total spectral
indices.  Note, in particular, that the objects with inverted core
spectral indices and steep total spectral indices in the upper left
hand side of Figure~\ref{fig:fig2} are all lobe dominated, which is
indicative of a resolution effect as opposed to a variability effect.
Therefore, it seems likely that true radio variability is not a
significant source of error in our previously published low resolution
spectral data.

It is also important to realize that variability affects higher
frequency data more than low frequency data, since the higher
frequency observations are more sensitive to flat-spectrum sources,
which are more likely to be variable.  A comparison of flux densities
of 20\,cm data from different epochs should reveal less variability
than a similar comparison at 6\,cm.  Therefore, the variability in
Figure~\ref{fig:fig1} is only a lower limit to the variability in the
spectral indices.

\begin{figure}[p]
\epsscale{1.0}
\plotone{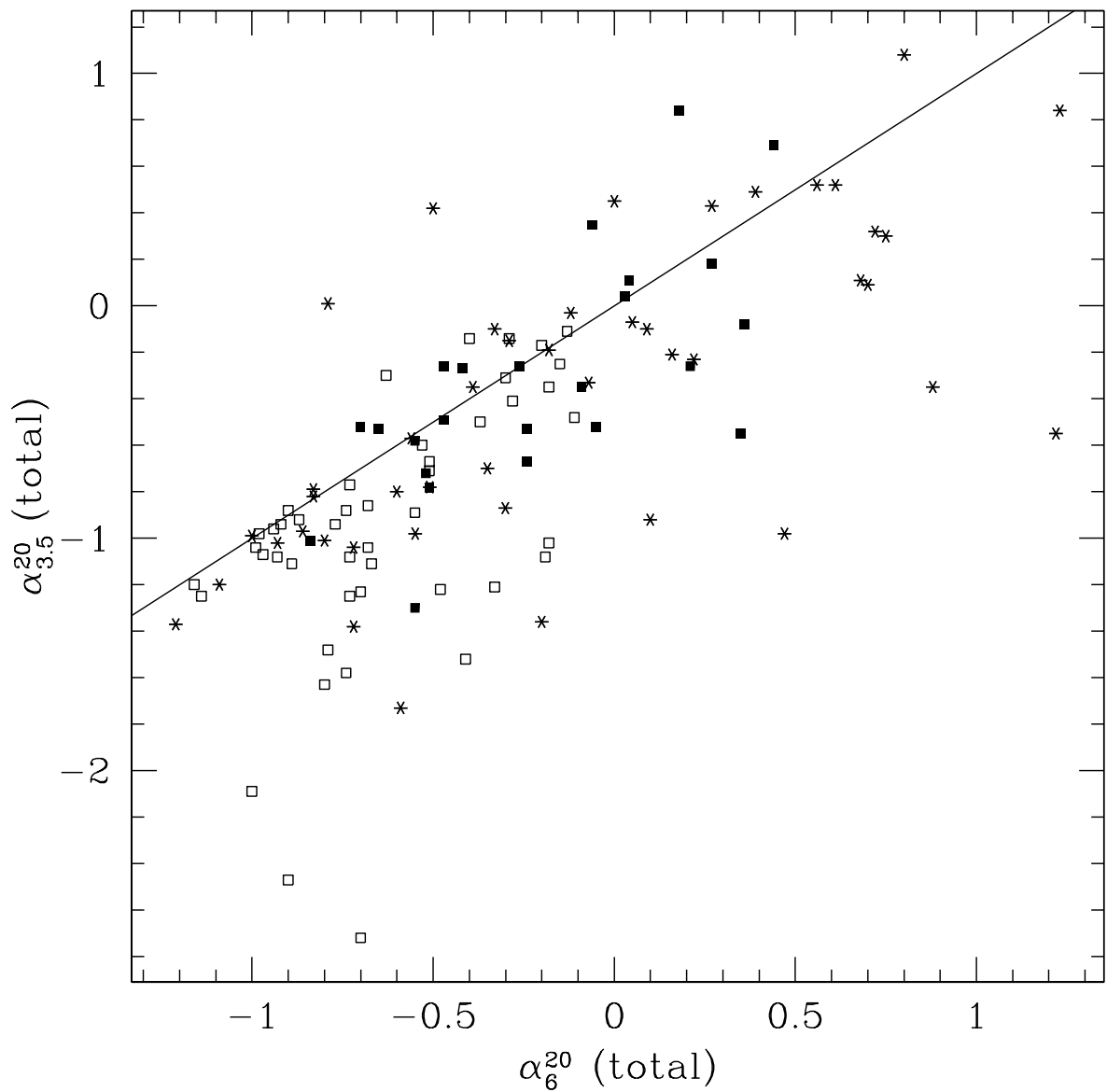}
\caption{Total (core+extended) spectral index versus total spectral
index.  Points are labeled as in Figure~\ref{fig:fig2}.  As in
Figure~\ref{fig:fig2}, the solid line traces the values where the
spectral indices are equal: it is not a fit to the
data. \label{fig:fig3}}
\end{figure}

Having addressed the effect that variability has on the measurement of
radio spectral indices, we now turn to an analysis of their use as
quasar orientation measures.  Our new data allows the computation of
spectral indices between 3.5 and 20\,cm (at high resolution), whereas
we used the 6 and 20\,cm flux densities (at low resolution) in our
previous analysis (Paper I).  As such, it is worth making a comparison
of the total (core+extended) spectral indices as determined from each
data set; Figure~\ref{fig:fig3} depicts this comparison.  The spectral
indices from the old and new data are roughly correlated; however, the
slope of the correlation is not unity and the majority of the points
are steeper in the high resolution data.

\placefigure{fig:fig3}

Nominally, one might expect that the core-dominated sources will have
flat spectral indices independent of resolution.  Similarly, the
spectral indices of lobe-dominated sources would be steep both at low
and high resolution, although less so in the A-configuration data
because over-resolution would cause one to miss a significant fraction
of the extended steep spectrum emission, thus flattening the overall
spectrum.  These naively expected trends are evident in the data,
though there are exceptions.  About a dozen of the core-dominated
sources exhibit steep radio spectra (as measured by either index).
These sources constitute a larger than expected fraction of
core-dominated steep spectrum sources ($\sim$ 40\% of the
core-dominated sources in Figures~\ref{fig:fig2} and~\ref{fig:fig3}
are steep-spectrum).  Some of these sources may be Compact Steep
Spectrum objects \citep[CSS;][]{ode98}, although our division of core-
vs. lobe-dominated (C$_{20}$ less than or greater than unity) is not
the same as is used to define CSS sources.

In addition to variability, we were also concerned that the resolution
of our old (Paper I) radio data might bias our results.  Since we
chose to make our new radio maps at higher resolution (using the A
configuration of the VLA), it is interesting to ask how much influence
resolution has on spectral indices, especially how resolution affects
the degree to which spectral indices correlate with the orientation
angle of AGN.  Figure~\ref{fig:fig2} shows how the core spectral index
(between the 3.5 and 20\,cm) correlates with the total spectral index
(between the 6 and 20\,cm).  There is a correlation, though it is not
as strong as in Figure~\ref{fig:fig3}.  Of particular interest are the
lobe-dominated sources with flat cores located in the middle of the
left hand side of the plot.  If spectral indices are indeed a measure
of the orientation of quasars, then it is clear that one or both are
failing for these objects.  For the most inverted-spectrum sources
(both spectral indices greater than $-0.4$) both the core and total
spectral indices agree much better.

\begin{figure}[p]
\epsscale{1.0}
\plotone{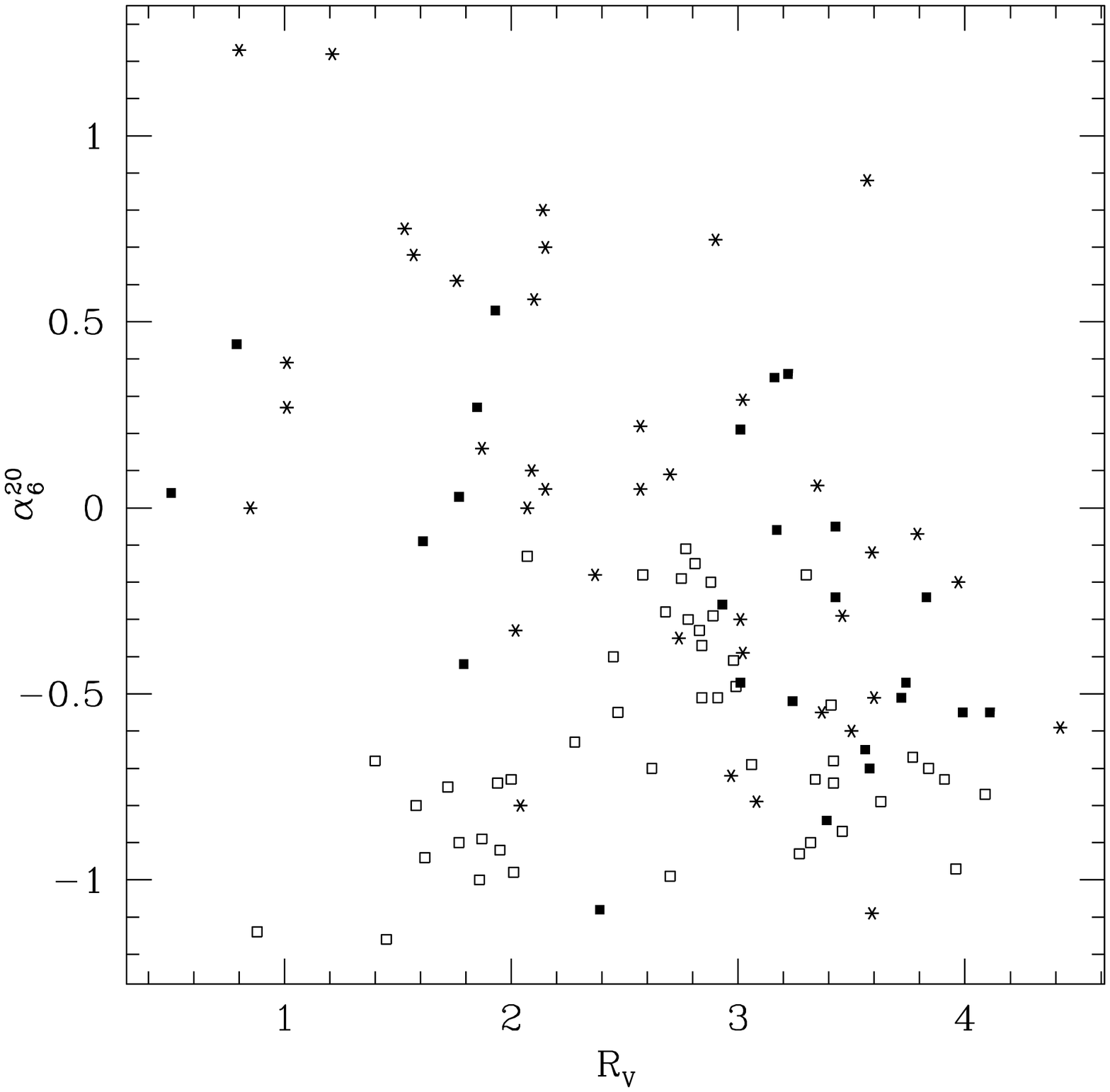}
\caption{Total spectral index versus the ratio of core radio
luminosity to optical luminosity.  Points are labeled as in
Figure~\ref{fig:fig2}. \label{fig:fig4}}
\end{figure}

Though radio spectral indices may be good orientation indicators,
\citet{wb95} claim that the ratio of core radio luminosity to optical
luminosity ($R_V$) may be a better indicator of quasar orientation,
particularly as compared to the ratio of core radio luminosity to
extended radio luminosity ($C$, the core-to-lobe ratio).  Their
argument is that the extended radio flux is sensitively dependent upon
interactions with the quasar environment, whereas the optical
luminosity is correlated with the power available to the lobes and is
not affected by the quasar environment.  In Figure~\ref{fig:fig4} we
compare the total (low resolution) spectral index to this parameter
($R_V$).  There is a reasonable correlation between the two parameters
if one ignores the lobe-dominated steep (total) spectrum sources that
have anomalously low values of $R_V$.  These are the same outliers as
in Figure~\ref{fig:fig2}.  If both parameters are orientation
measures, then one is clearly failing on these sources.  Given that
these objects have lobe-dominated morphologies and have total spectral
indices that are consistent with this observation, it would seem that
$R_V$ is the parameter in ``error''.

\placefigure{fig:fig4}

\begin{figure}[p]
\epsscale{1.0}
\plotone{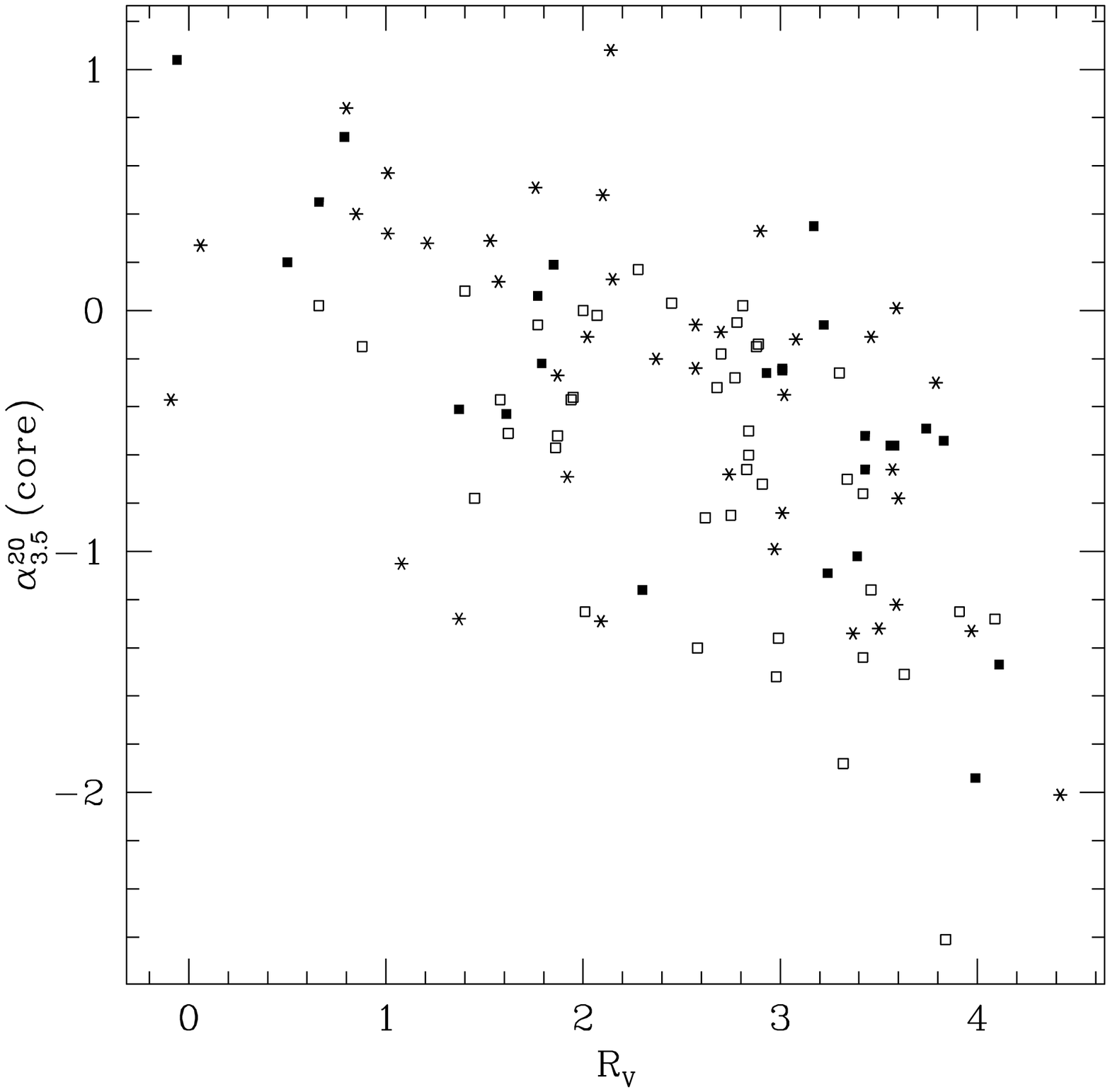}
\caption{Core spectral index versus the ratio of core radio luminosity
to optical luminosity.  Points are labeled as in
Figure~\ref{fig:fig2}. \label{fig:fig5}}
\end{figure}

For comparison, we also plot $R_V$ versus the core spectral index in
Figure~\ref{fig:fig5}.  Both of these parameters are ratios that
contain the core 20\,cm flux, so the correlation between the two is
somewhat artificial.  Nevertheless, we can see that objects with flat
cores tend to have small values of $R_V$, whereas objects with steep
cores tend to have larger values of $R_V$.

\placefigure{fig:fig5}

\begin{figure}[p]
\epsscale{1.0}
\plotone{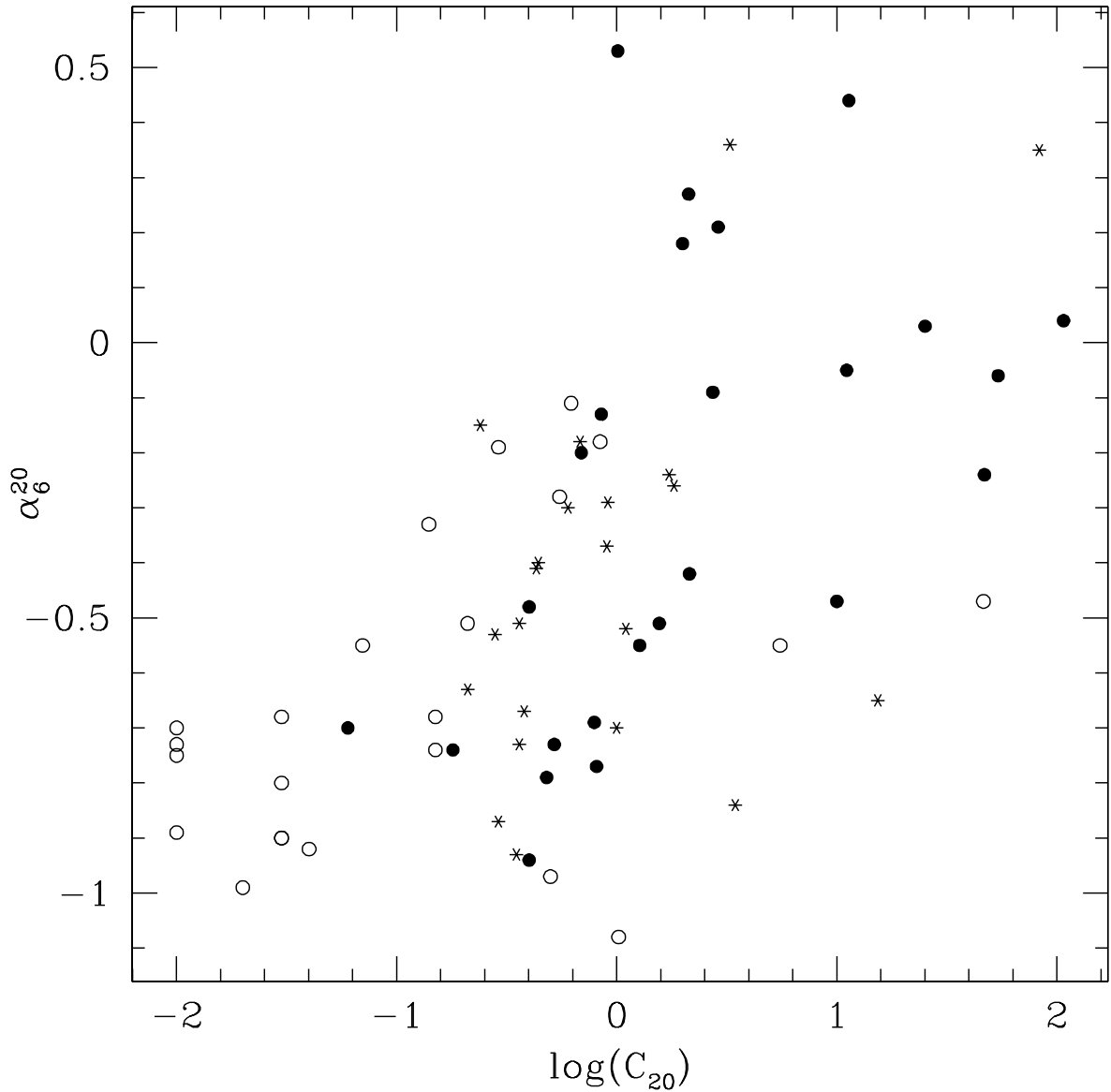}
\caption{Total spectral index versus the log of the core-to-lobe
ratio.  Points are labelled according to morphology of their 20\,cm
maps.  Open circles are those objects whose 20\,cm maps are resolved
into two or more sources.  Closed circles are apparently unresolved
point sources.  Stars are objects that were difficult to classify as
one or the other.  Note that the morphologies are by eye
classifications and are meant to be illustrative only and not
quantitative. \label{fig:fig6}}
\end{figure}

As a final test, we compare the total (low resolution) spectral
indices against the core-to-lobe ratio (at 20\,cm).
Figure~\ref{fig:fig6} gives the log of the core-to-lobe ratio versus
the total (low resolution) spectral index.  Open circles are those
objects whose 20\,cm maps are resolved into two or more sources.
Closed circles are apparently unresolved point sources.  Stars are
objects that were difficult to classify as one or the other.  There is
a general (but not perfect) trend towards flatter spectral indices
with more core-dominated sources.  The most lobe-dominated sources
($C_{20}<0.1$) are steep-spectrum sources ($\alpha < -0.4$), as
expected.

\placefigure{fig:fig6}

One caveat with respect to Figure~\ref{fig:fig6} is that the
core-to-lobe ratio is best used as an indicator of relative
orientation only among classes of objects that are believed to be
intrinsically the same.  Because this sample was chosen based on the
presence of optical absorption properties, there is little reason to
believe the radio engines are intrinsically the same in these objects.
Specifically, the bulk jet Lorentz factors may vary significantly from
object to object.  This fact will adversely affect the core-to-lobe
ratio as an orientation indicator since it implies two objects that
are intrinsically the same in every respect except in jet speed, seen
at identical lines of sight, will have significantly different
core-to-lobe ratios.  Finally, the fact that different classes of
radio sources have different {\em intrinsic\/} amounts of core
relative to lobe emission, may introduce a scatter into
Figure~\ref{fig:fig6}.  Core-to-lobe ratios are therefore best used as
orientation measures when there is some reason to believe the radio
sources in question are intrinsically the same and only appear
different because of orientation.  Again, this is not necessarily the
case for this sample of quasars, selected by virtue of the presence of
optical \ion{C}{4} absorption.

From an analysis of our new radio data in comparison to the radio data
used in Paper I, we have addressed the usefulness of various radio
properties as quasar orientation measures.  Given the rough
correlation between each of these properties, it is likely that all
trace the orientation of quasars to some degree.  However, it is clear
that the correlation between these parameters is far from perfect and
that these ``orientation indicators'' should be treated with a healthy
amount of caution.  

The primary purpose of using multiple orientation indicators in this
study was to test the validity of the total flux radio spectral indices
as an orientation indicator in Papers I and II.  That the total flux
radio spectral index is roughly correlated with the other properties
presented herein is taken as justification for the use of this
parameter.  Even if the radio spectral indices are not the best
measure of orientation per se, there is no doubt that they are probes
of properties that are intrinsic to the quasars and therefore should
not correlate in any way with any property {\em not\/} intrinsic to
the quasar, such as intervening absorption systems.  We now turn our
analysis to this issue.

\section{Absorption Line Analysis and Discussion}

The purpose of obtaining this new radio data was to determine if our
inhomogeneous radio data were biasing the results of our analysis from
Paper I.  At least three issues are involved here.  First, the
spectral indices in Paper I were determined from maps made at
different epochs.  Second, the old radio data are low resolution.
Finally, only the brightest sources had spectral index measurements.
With the new radio data presented above, we have alleviated each of
these problems and we can now proceed with a re-analysis of the
absorption line properties of quasars as a function of quasar
orientation measures.

The new data seem to indicate that, but for a small number of sources,
the total spectral indices are reasonably correlated with other
intrinsic quasar properties that have traditionally been used as
orientation measures.  In particular, the total spectral indices taken
nearly simultaneously and at high resolution agree well with the total
spectral indices taken at different epochs and at low resolution.
This being the case, we could stop here and claim that our results
support the conclusions drawn in Paper I.  However, it is worth
discussing this and related issues in further detail.

The absorption line data is studied as a function of the intrinsic
radio properties of six subsamples of data.  The results from this
analysis are presented in Figure~\ref{fig:fig7} and
Table~\ref{tab:tab3}.  The samples are defined so as to allow for the
splitting of the samples into two based on the orientation indicator
that defines the sample.  Sample A includes the \ion{C}{4} absorbers
in quasars where the radio spectral index has been measured at low
resolution between 6 and 20\,cm.  Sample B is a subset of sample A and
includes only those quasars for which we have both low resolution
spectral indices between 6 and 20\,cm and also high resolution, core
spectral indices between 3.5 and 20\,cm.  Sample C includes only those
quasars for which we were able to measure high resolution, core
spectral indices between 3.5 and 20\,cm.  Sample D combined both
samples B and C by requiring not only that both high and low
resolution spectral indices are measured, but also that both spectral
indices are both either steep or flat.  Sample E includes those
objects for which we measured $R_V$, whereas Sample F includes the
quasars that have measured values of $C_{20}$.  The columns in
Table~\ref{tab:tab3} are as follows.  Column 1 gives the sample and
the orientation indicator (e.g. steep or flat spectrum).  Columns 2-4
give values for absorbers with velocities within $\pm 5000\,{\rm
km\,s^{-1}}$ of the quasar redshift, including 1) $dN/d\beta$, 2) the
error in $dN/d\beta$, and 3) the number of absorbers.  Columns 5-8
give values for absorbers with velocities larger than $5000\,{\rm
km\,s^{-1}}$, but less than $55,000\,{\rm km\,s^{-1}}$, including 5)
$dN/d\beta$, 6) the error in $dN/d\beta$, 7) the observed number of
absorbers, and 8) the expected number of absorbers based on the
average of the number of quasars studied.

\placefigure{fig:fig7}
\placetable{tab:tab3}

\subsection{Narrow, High-Velocity \ion{C}{4} Absorption}

\begin{figure}[p]
\epsscale{1.0}
\plotone{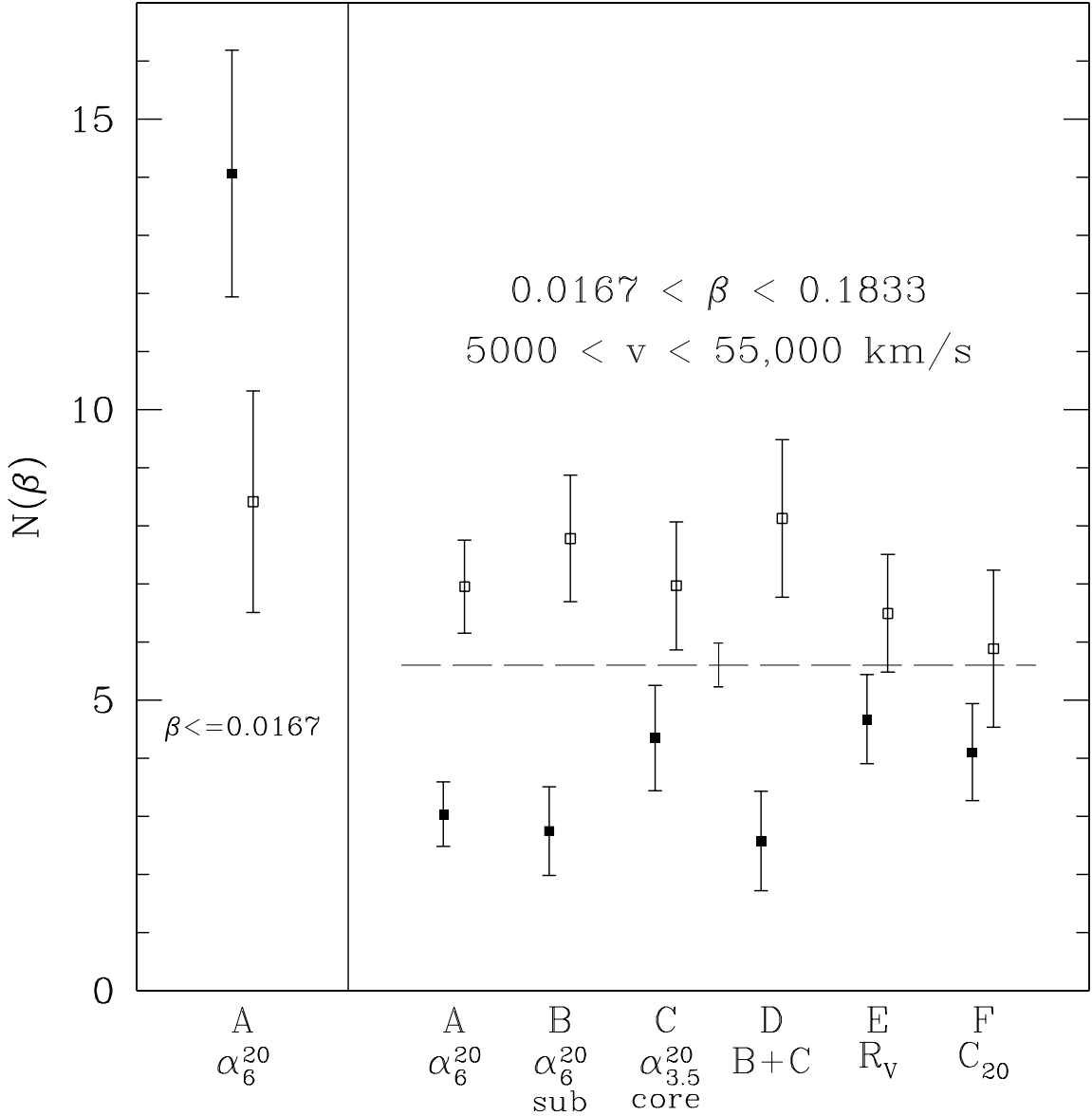}
\caption{Velocity distribution of \ion{C}{4} absorbers for several QSO
samples with different intrinsic radio properties.  Two velocity
ranges are displayed.  On the left we show the velocity distribution
for all absorbers within $\pm 5000\,{\rm km s^{-1}}$ with respect to
the quasar redshift for quasars with steep (solid square) and flat
(open square) orientation indicators.  On the right, we show the
velocity distribution of \ion{C}{4} with
$\beta=(v/c)>0.0167\,(v>5000\,{\rm km s^{-1}}$) for six samples of
QSOs, including the sample for which we plot the low velocity systems.
The dashed line shows the average value for all quasars (Paper II).
The errors given are $1\sigma$ Poisson errors.  \label{fig:fig7}}
\end{figure}

In Paper I we showed that there is a excess of high-velocity
\ion{C}{4} absorbers in flat-spectrum quasars as compared to
steep-spectrum quasars.  We also showed that this effect is not
apparently the result of some other bias in the steep-spectrum quasar
sample (i.e. because they tend to be fainter or at lower redshift).
In Figure~\ref{fig:fig7} we show the apparent velocity distribution of
\ion{C}{4} absorbers for a number of samples of QSOs with different
observed radio properties.  Here we assume that the redshifts of the
absorbers are due to outflows from QSOs and are non-cosmological (we
certainly do not propose that {\em all} \ion{C}{4} absorption is
non-cosmological; however, in making this plot that is a fundamental
assumption).

For sample A, we plot the number density of absorbers per unit
velocity versus velocity for both steep- and flat-spectrum quasars.
The spectral indices are the same as those used in Paper I,
specifically the ratio of the 6 and 20 cm flux densities from Green
Bank (or NVSS).  A spectral index of -0.4 ($f_{\nu} \propto
\nu^{\alpha}$), which is the median of this sample (as compared to
-0.5, used in Paper I), serves as the dividing line between flat- and
steep-spectrum sources.  For sample A, the closed squares are the
steep-spectrum data points, whereas the open squares are the
flat-spectrum data points.  On the far left of Figure~\ref{fig:fig7},
we include only those absorbers within $\pm 5000\,{\rm km\,s^{-1}}$ of
the QSO redshift (sample A only); these are the so-called
``associated'' absorption systems \citep[e.g.,][]{fwp+86}.  For the
remainder of the samples only higher velocity data points are given,
where the low velocity edge of the bin is at $5000\,{\rm km\,s^{-1}}$,
and the red edge is at $55,000\,{\rm km\,s^{-1}}$.  This range is set
so as to maximize the number of absorbers in the bin, but to avoid
lines that are either too close to the QSO redshift or that are too
close to the Lyman-$\alpha$ emission line.  Except for the binning of
the high velocity data (which is necessary for comparison for the
samples that follow, since they have fewer absorption line systems),
sample A represents a compression of the data in the top panel of
Figure 4 in Paper I.

As was observed in Paper I there is an excess of high velocity
\ion{C}{4} absorption between $5000$ and $55,000\,{\rm km\,s^{-1}}$ in
flat-spectrum quasars as compared to steep-spectrum quasars.  While
the significance of this excess is not large and the numbers are
small, the potential consequences for quasar absorption line studies
justifies further analysis in attempt to confirm or reject this
result.  In addition to the tests presented herein and in Papers I and
II, it would be extremely useful to conduct a similar study that
focuses on bright steep-spectrum quasars since they tend to be fainter
than flat-spectrum quasars \citep{ob82}, even though such a bias was
ruled out in Paper I.

The first test we conduct using our new radio data is to ask if this
result could be biased by variability of the radio measurements.
Using only those quasars from sample A for which we have core spectral
indices from our VLA A configuration data at 3.5 and 20\,cm, we repeat
the above experiment.  This comparison is necessary to test that our
selection of objects for followup radio observations is not biased
with respect to the initial sample.  Thus, in sample B (a subset of
sample A) we show the difference in the number density of \ion{C}{4}
absorbers between steep- (closed square) and flat-spectrum (open
square) quasars, again using the low resolution spectral indices.
That the steep/flat dichotomy is as strong or stronger in sample B as
in sample A supports the notion that this dichotomy is neither due to
variability or to a selection bias in the radio data.

Though there may be resolution and bandpass issues (our new high
frequency datapoint is at 3.5 cm, instead of 6 cm) between the high
resolution total spectral index and total spectral index measured at
lower resolution, the new spectral indices are unaffected by source
variability and were shown to correlate very well with the indices
derived from the low resolution data (Figure~\ref{fig:fig3}).  Thus,
it would seem then that our initial result from Paper I is probably
{\em not} the result of variability in the original radio data.

Since the results from both total spectral index measurements
(non-simultaneous low resolution and simultaneous high resolution)
agree well, we next look at the distribution of absorbers with respect
to the core spectral index.  If the core spectral index is a better
orientation indicator than the total spectral index and if there is an
orientation dependent population of absorbers, then we expect that
this experiment will yield a more dramatic steep/flat dichotomy.
Doing this experiment, we find that there is a difference in the
velocity distribution of \ion{C}{4} absorbers between flat- and
steep-spectrum sources when using core spectral indices; however, the
difference is weaker than expected from our hypothesis.  That this is
the case can be seen in sample C of Figure~\ref{fig:fig7}.

One possible explanation for the observation that core spectral
indices produce less of a dichotomy than total spectral indices is if
there is a population of high-velocity intrinsic \ion{C}{4} absorbers,
it could be located on scales much larger than the core radio flux.
For the 20\,cm data, this means that the absorbing material would have
to be $\sim2\,{\rm kpc}$ or more from the central engine of the
quasar, where
\begin{equation}
d = \frac{2c}{H_{\rm o}}\frac{\left[(1+z)-(1+z)^{1/2}\right]}{(1+z)^2}\theta,
\end{equation}
taking $q_{\rm o}=0.5$, $H_{\rm o} = 65\,{\rm km/s/Mpc}$,
$z_{em}=2.5$, and $\theta = 0.7\arcsec$.  If there are indeed narrow,
ejected absorbers, such a distance may not be surprising.  A velocity
of $1000\,{\rm km\,s^{-1}}$ corresponds to $10^{-3}\,{\rm
pc\,yr^{-1}}$, such that material might be found out to a distance of
$100\,{\rm kpc}$ during the expected lifetime of a quasar ($\sim
10^8\,{\rm yrs}$).  Alternatively, it could be that for the sample of
quasars studied here (selected originally by the presence of C IV
absorption systems), core radio spectral indices are not as accurate a
measure of either intrinsic or orientation-dependent differences as
the spectral indices measured using the total flux.

One way to test these conclusions further is to require that {\em
both} the low and high-resolution spectral indices be steep or flat
(i.e. combine samples B and C).  Cutting the data in this way has two
effects.  First, we are still using the low resolution spectral
indices as an orientation indicator.  Second, we remove any objects
that might be affected by variability by requiring that the high
resolution spectral index (which is derived from simultaneous data) be
the ``same'' (either steep or flat spectra) as the low resolution
spectral index.  Sample D in Figure~\ref{fig:fig7} shows the results
of such an analysis.  Even though the observed number of absorbers is
quite small in this subsample, it is again clear that there are more
absorbers in flat-spectrum quasars and fewer absorbers in
steep-spectrum quasars as compared to each other (and to the expected,
average, distribution).

A few comments regarding the distribution of \ion{C}{4} absorbers as a
function of other orientation indicators ($R_V, C_{20}$) are
necessary.  Looking at Figure~\ref{fig:fig4}, we note that to divide
the sample into two by the median of $\alpha_6^{20}$ produces
completely different data sets than if the sample is divided by the
median of $R_V$.  It is worth asking if the \ion{C}{4} absorption
dichotomy remains if the sample is delineated by ``$R_V$''.  In fact,
it does not.  High-$R_V$ and low-$R_V$ QSOs have nearly the same
velocity distribution of \ion{C}{4} absorbers (see Sample E in
Figure~\ref{fig:fig7}).  Since $R_V$ is thought to be one of the
better orientation measures, our hypothesis that there is a population
of intrinsic \ion{C}{4} absorption line systems that are a function of
QSO orientation seems to be in jeopardy.  However, we reiterate our
doubt that $R_V$ is a good orientation measure for our particular
sample, since the $R_V$ values for lobe-dominated, steep-spectrum
sources are clearly out of line with the expected values.

There is a similar effect with the core-to-lobe ratios, $C_{20}$.
From Figure~\ref{fig:fig6} we see that whereas lobe-dominated quasars
($C_{20} \le 1$) are predominantly steep-spectrum sources, and
core-dominated quasars ($C_{20} > 1$) are typically flat-spectrum
sources, there are exceptions.  As with $R_V$, a division of the
sample into two parts using $C_{20} = 1$ as the dividing line yields
little or no difference in the \ion{C}{4} absorber distribution
between the two subsamples of sample F in Figure~\ref{fig:fig7}.  This
effect would also seem to rule against the conclusion that many high
velocity \ion{C}{4} absorbers are actually intrinsic to the quasars.
However, the range of spectral indices for core-dominated objects is
large.  There is a handful of core-dominated steep-spectrum sources
and also a number of objects that are multiple or resolved to the eye,
but have core-to-lobe ratios that suggest that they are
core-dominated.  Furthermore, both lobe- and core-dominated
steep-spectrum sources show a dearth of absorbers.  These issues need
to be resolved in order for an analysis of the distribution of
\ion{C}{4} absorbers as a function of core-to-lobe ratio to be
meaningful.

While some measures of properties intrinsic to quasars (e.g., $C_{20}$
and $R_V$) do not show a statistically significant difference in the
frequency of high velocity \ion{C}{4} absorbers, other indicators
(e.g., various estimates of the radio spectral index) clearly show a
statistical difference between flat- and steep-spectrum objects.  This
could be interpreted several ways.  First, detection of high velocity
\ion{C}{4} absorbers could correlate with quasar orientation (and
therefore be intrinsic to the quasars themselves) and two common
indicators of quasar orientation ($R_V$ and $C_{20}$) are simply not
good orientation measures for this particular quasar sample for any
number of reasons.  Alternatively, it could be that the presence of
high velocity \ion{C}{4} absorbers is {\em not} related to quasar
orientation in any simple way, but {\em is} related to at least one
intrinsic quasar property: the total (core+extended) radio spectral
index.  Either way, our main conclusion from Paper I is confirmed
here: the frequency of narrow, high-velocity \ion{C}{4} absorption
systems cannot all be due to intervening systems.  At least some
fraction of these systems must be intrinsic to the quasars themselves.

\subsection{``Associated'' Absorption}

In the left panel of Figure~\ref{fig:fig7}, sample A represents those
absorption lines that have redshifts within $5000\,{\rm km\,s^{-1}}$
of the QSO emission redshift.  These absorption lines are the
so-called ``associated'' absorption systems \citep{fwp+86}.  Clearly,
there is little doubt that these systems are drawn from a population
that is different from the systems observed at larger relative
velocities.  We confirm that these systems are atypically strong and
tend towards steep-spectrum quasars.  We also emphasize (as we did in
Paper I) that the excess of low velocity absorbers in steep-spectrum
quasars is made more evident by the dearth of high-velocity absorbers
in the same.  On the other hand, the flat-spectrum level appears to be
more consistent as a function of velocity
(c.f. Figure~\ref{fig:fig7}).

Given that the so-called associated absorption systems tend to be
``associated'' with steep-spectrum quasars, it might be interesting to
look at the other orientation measures for QSOs with associated
absorption.  We find that the average $\alpha_6^{20}$ (total, low
resolution) for QSOs from Paper I with \ion{C}{4} absorption within
$3000\,{\rm km\,s^{-1}}$ of the emission redshift is $-0.76$.
Unfortunately, our new sample only included a half-dozen or so of
these quasars, so it is difficult to do a complete analysis of the
other orientation measures; however, we can calculate the mean values.
We find that the average associated absorption quasar has
$\alpha_{3.5}^{20}$ (core, high resolution) $= -0.49$,
$\alpha_{3.5}^{20}$ (total, high resolution) $= -1.07$, $R_V = 2.55$,
$\alpha_6^{20}$ (total, low resolution) $= -0.82$, while the $C_{20}$
values are intermediate.  Curiously, both of the ``total'' spectral
indices (high and low resolution) are very steep, whereas the core,
high resolution spectral index is near the dividing line.  $R_V$ and
$C_{20}$ are also right near the dividing line.  

It is interesting that the associated systems are found so
predominantly in steep-spectrum sources, yet their other orientation
measures are more intermediate.  This situation is similar to what we
observe for the narrow, high-velocity absorbers: the excess of
absorbers only occurs for comparisons of total spectral indices, but
not other orientation measures.  That this behavior occurs in a sample
of systems that are accepted as being associated and not intervening
lends credence to the possibility that there may be a population of
narrow, high-velocity \ion{C}{4} absorbers that are related to the QSO
and are distributed as a function of the total radio spectral index.
Further work is clearly needed to fully understand the associated
absorbers; however, a detailed analysis is beyond the scope of this
work.

\subsection{Orientation versus Intrinsic Indicators}

Throughout this work we refer to a number of radio properties of
quasars as ``orientation indicators''.  The reader may agree or
disagree that these measurements do indeed correlate with the
line-of-sight orientation of quasars in a statistical sample.
However, it is important to stress that it is not very important to
this work that these parameters be good orientation indicators.  All
that is required is that they are good indicators of {\em something}
intrinsic to the quasar itself.  That is, the radio properties are all
intrinsic to the quasars even if they are not good orientation
measures; the fact that these properties are intrinsic and not that
they are orientation indicators are what make them ideal for studying
intrinsic absorption in quasars.  For example, the excess of
\ion{C}{4} absorbers at low velocities in steep-spectrum quasars is an
empirical fact.  Whether or not the spectral indices are also a good
orientation indicator is not entirely relevant; the total radio
spectral index need not delineate the location of the jet or the axis
of the disk so long as it tells us something about the physical
location and distribution of the hypothesized absorbing material.
However, it {\em is} true that in terms of producing a model that
explains such effects, it would be surprising if the spectral index
was not a good orientation indicator.  As such, our discussion leans
towards the assumption that these intrinsic radio properties of
quasars are indicators of the orientation of the quasars.

In terms of intrinsic absorption, whether or not the spectral indices
are also orientation indicators depends significantly on the location
of the absorbing material with respect to the material that is
producing the radio emission.  Since quasars are observed to have
extended radio emission, it is possible that the absorbing material
could be located anywhere from about 1\,pc (the scale of the accretion
disk) to hundreds of kiloparsecs (the scale of a large radio lobe).
If intrinsic absorbers, whether they be the accepted type, such as the
``associated'' absorbers, or the high-velocity type postulated herein,
are formed far from the central engine of the quasar, such as at the
boundary where the radio lobes interact with the IGM, then it would
not be surprising to find that there is a correlation between
intrinsic absorption properties and total spectral index, but not
quasar orientation.  This is because the spectral index will have been
profoundly influenced by the IGM material (which is why \citet{wb95}
proposed using $R_V$ as an orientation measure in the first place),
and also because precession of the central source during the lifetime
of the lobes may produce a significant offset.  As such, we too
caution against the blind used of radio properties as orientation
measures, but we stress that their use as such is not explicitly
required for our analysis.

\section{Conclusions}

We have presented new radio measurements at both 3.5 and 20\,cm using
the VLA in the A configuration for 144 quasars with known \ion{C}{4}
absorption line properties.  These measurements are used to determine
radio properties that are often used as quasar orientation measures,
particularly the total radio spectral index as determined from
non-simultaneous data in two bandpasses.

Our results can be summarized as follows:

1) Total, low-resolution spectral indices (between 6 and 20\,cm)
   computed from data taken at different epochs are not significantly
   corrupted by variability.  Furthermore, they are reasonably well
   correlated with total spectral indices computed between 3.5 and
   20\,cm that were taken simultaneously at much higher resolution.

2) $R_V$ is correlated with $\alpha_6^{20}$ (total, low resolution),
   except for some lobe-dominated sources, which appear to have
   ``incorrect'' values of $R_V$.

3) We confirm the excess of narrow, high-velocity \ion{C}{4} absorbers
   found in flat-spectrum quasars as compared to steep-spectrum
   quasars that was reported by \citet{ryy+99}.  This excess is not
   caused by variability or a radio luminosity bias.

4) The steep/flat dichotomy is less pronounced for core spectral
   indices.  If there is indeed a population of narrow, high-velocity
   intrinsic absorbers, this fact may indicate that the location of
   these systems is beyond the extent of the resolution of the VLA in
   the A configuration ($\sim 2\,{\rm kpc}$).

5) Other orientation measures do not show the same dichotomy.
   However, there is reason to believe that these measures are flawed
   as good orientation indicators for our particular sample.
   Furthermore, the dichotomy need not be strictly an orientation
   effect, but rather an intrinsic effect.

6) Associated absorption systems are preferentially found in quasars
   with total spectral indices that are very steep.  However, the mean
   values of other orientation measures are more moderate.

\acknowledgements 

RHB and SLM acknowledge support from the NSF (grants AST-98-02791 and
AST-98-02732) and the Institute of Geophysics and Planetary Physics
(operated under the auspices of the U.S. Department of Energy by the
University of California Lawrence Livermore National Laboratory under
contract No. W-7405-Eng-48).  We thank an anonymous referee for
suggestions that improved the paper.

\begin{deluxetable}{lrrrrrrrrrrr}
\tabletypesize{\small}
\rotate
\tablewidth{0pt}
\tablecaption{Observed Quasar Properties \label{tab:tab1}}
\tablehead{
\colhead{Name} & 
\colhead{RA} & 
\colhead{Dec} &
\colhead{$z_{em}$} & 
\colhead{mag} &
\colhead{6\,cm} &
\colhead{20\,cm} &
\colhead{peak} &
\colhead{total} &
\colhead{peak} &
\colhead{total} \\
\colhead{} & 
\colhead{(J2000)} & 
\colhead{(J2000)} &
\colhead{} & 
\colhead{} &
\colhead{} &
\colhead{} &
\colhead{3.5\,cm} &
\colhead{3.5\,cm} &
\colhead{20\,cm} &
\colhead{20\,cm}
}
\startdata
FBQS0047-0156 &  0:47:17.80 &  -1:56:23.85 & 2.48 & 17.9 &    \nodata &    3.7 &    3.03 &    2.92 &    \nodata &    \nodata \\ 
FBQS0048-0103  &  0:48:06.08 &  -1:03:21.52 & 2.53 & 18.3 &    \nodata &    2.3 &    3.80 &    3.69 &    \nodata &    \nodata \\ 
FBQS0210-0152 &  2:10:39.88 &  -1:52:13.81 & 2.37 & 18.3 &    \nodata &    1.8 &    \nodata &    \nodata &    \nodata &    \nodata \\ 
FBQS0256-0119 &  2:56:25.63 &  -1:19:11.91 & 2.49 & 18.2 &    \nodata &   22.9 &    6.96 &    8.18 &   23.81 &   28.10 \\ 
FBQS0729+2524 &  7:29:28.47 &  25:24:51.80 & 2.30 & 17.7 &   45.0 &    9.9 &   37.60 &   38.57 &   22.91 &  101.84 \\ 
FBQS0804+2516 &  8:04:13.67 &  25:16:33.63 & 2.30 & 17.8 &    \nodata &    4.8 &    0.46 &    0.95 &    3.03 &    7.72 \\
FBQS0857+3313 &  8:57:26.93 &  33:13:16.86 & 2.34 & 17.5 &   30.0 &   21.4 &   33.29 &   34.78 &   18.78 &   16.09 \\ 
FBQS0934+3153  &  9:34:04.00 &  31:53:31.17 & 2.42 & 17.9 &    \nodata &    6.5 &    4.33 &    4.15 &    4.16 &    7.17 \\ 
FBQS0955+3335  &  9:55:37.94 &  33:35:03.95 & 2.50 & 17.0 &   55.0 &   33.8 &  114.54 &  115.16 &   41.47 &   47.83 \\ 
FBQS1045+3440 & 10:45:41.03 &  34:40:43.38 & 2.35 & 17.9 &   19.0 &   21.2 &    9.24 &   10.58 &   19.90 &   19.79 \\ 
FBQS1253+2905  & 12:53:06.42 &  29:05:14.14 & 2.56 & 17.5 &   40.0 &   67.8 &   41.64 &   42.18 &   61.68 &   68.65 \\ 
FBQS1348+2840  & 13:48:04.34 &  28:40:25.50 & 2.47 & 17.6 &  105.0 &   86.2 &   37.59 &   38.69 &   60.50 &   56.60 \\ 
FBQS1457+2707  & 14:57:05.23 &  27:07:57.28 & 2.53 & 17.4 &   32.0 &    2.7 &    5.04 &    6.21 &    \nodata &    \nodata \\ 
FBQS1537+2716 & 15:37:14.26 &  27:16:11.52 & 2.43 & 17.2 &    \nodata &    1.6 &    4.17 &    5.09 &    2.59 &    4.31 \\ 
FBQS1540+4138 & 15:40:42.98 &  41:38:16.40 & 2.51 & 17.1 &   20.0 &   19.9 &   42.04 &   43.80 &   20.63 &   19.78 \\ 
FBQS1625+2646 & 16:25:48.78 &  26:46:58.51 & 2.52 & 17.2 &    \nodata &    6.4 &    2.43 &    2.91 &    \nodata &    \nodata \\ 
FBQS1634+3203 & 16:34:12.79 &  32:03:35.27 & 2.33 & 17.1 &  248.0 &  177.2 &  209.62 &  208.71 &  150.50 &  150.20 \\ 
FBQS1651+4002 & 16:51:37.55 &  40:02:19.09 & 2.33 & 17.1 &   49.0 &   43.5 &    4.75 &    9.01 &   47.11 &   46.61 \\ 
FBQS2216-0057  & 22:16:08.88 &   0:57:08.28 & 2.39 & 18.0 &    \nodata &    7.4 &   14.01 &   15.37 &    6.33 &    7.70 \\ 
FBQS2233-0838 & 22:33:38.72 &  -8:38:53.00 & 2.34 & 18.2 &    \nodata &    9.3 &    4.42 &    4.70 &    9.24 &   10.39 \\ 
0002+051       &  0:05:20.18 &   5:24:10.90 & 1.90 & 16.2 &  296.0 &  127.2 &  182.39 &  183.29 &  146.92 &  152.00 \\ 
0004+171       &  0:06:47.27 &  17:28:15.70 & 2.90 & 18.5 &  159.0 &  226.3 &   98.81 &  100.35 &  173.97 &  206.99 \\ 
0014+813       &  0:17: 8.10 &  81:35: 7.80 & 3.39 & 16.5 &  551.0 &  692.9 &  489.94 &  497.06 &  696.62 &  698.93 \\ 
0017+154       &  0:20:25.31 &  15:40:53.31 & 2.02 & 18.2 &  486.0 & 2180.5 &    7.53 &  171.64 &  754.37 & 1981.30 \\ 
0033+0951      &  0:36:23.83 &  10:07:59.20 & 1.91 & 17.5 &  328.0 &  719.2 &  328.11 &  399.86 &  242.08 &  679.66 \\ 
0038-019       &  0:41:26.01 &  -1:43:16.00 & 1.67 & 18.5 &  416.0 & 1034.7 &   38.90 &  130.79 &   38.71 &  902.41 \\ 
0051+291       &  0:53:44.37 &  29:25: 7.00 & 1.83 & 17.8 &  255.0 &  510.3 &  141.12 &  169.24 &  173.75 &  469.14 \\ 
0109+176       &  1:11:49.87 &  17:53:51.00 & 2.15 & 18.0 &  160.0 &  488.4 &   35.46 &   88.78 &   39.58 &  425.56 \\ 
0119-046       &  1:22:27.91 &  -4:21:27.30 & 1.97 & 16.5 &  780.0 & 1473.4 &  340.87 &  414.72 &  994.64 & 1378.20 \\ 
0123+257       &  1:26:42.81 &  25:59: 1.50 & 2.36 & 17.5 & 1303.0 &  998.2 &  607.55 &  618.36 &  926.45 &  983.19 \\ 
0136+1737      &  1:39:41.86 &  17:53: 8.10 & 2.72 & 18.5 &  298.0 &  428.1 &  268.45 &  270.83 &  343.75 &  346.75 \\ 
0149+336       &  1:52:34.54 &  33:50:32.40 & 2.43 & 18.5 &  611.0 &  740.4 &  342.75 &  369.33 &  329.55 &  578.12 \\ 
0150-202       &  1:52:27.31 & -20:01: 7.14 & 2.15 & 17.1 &  122.0 &   26.3 &  120.27 &  121.37 &   27.04 &   27.22 \\ 
0201+3634      &  2:04:55.50 &  36:49:18.00 & 2.91 & 17.5 &  358.0 &  564.5 &  203.03 &  205.21 &  493.58 &  503.19 \\ 
0225-014       &  2:28: 7.83 &  -1:15:40.60 & 2.04 & 18.1 &  230.0 &  455.6 &   70.73 &   81.99 &  115.29 &  401.30 \\ 
0226-038       &  2:28:53.26 &  -3:37:38.20 & 2.07 & 17.0 &  570.0 &  933.0 &  554.41 &  630.90 &  528.35 &  805.65 \\ 
0229+1309      &  2:31:45.72 &  13:22:54.90 & 2.07 & 17.7 & 2427.0 & 1560.3 & 1289.50 & 1333.30 & 1426.90 & 1542.60 \\ 
0237-233       &  2:40: 8.11 & -23:09:18.00 & 2.23 & 16.6 & 3160.0 & 6256.9 &  440.24 &  600.14 & 5993.80 & 6051.00 \\ 
0238+100       &  2:41:22.19 &  10:18:47.00 & 1.83 & 18.0 &   87.0 &  272.6 &   21.23 &   48.46 &   39.97 &  257.25 \\ 
0256-000       &  2:59: 5.63 &   0:11:22.00 & 3.37 & 18.7 &    4.0 &    3.2 &   13.40 &   14.77 &    2.08 &    3.31 \\ 
0316-203       &  3:18:25.23 & -20:12:18.80 & 2.87 & 19.5 &   40.0 &  100.8 &    6.39 &    5.59 &   83.60 &   92.84 \\ 
0329-2534      &  3:31: 8.95 & -25:24:43.30 & 2.69 & 17.5 &  253.0 &  320.9 &   54.28 &   41.98 &  245.23 &  287.05 \\ 
0336-0142      &  3:39: 0.80 &  -1:33:18.00 & 3.20 & 19.1 &  475.0 &  592.9 &  299.70 &  303.03 &  478.07 &  564.59 \\ 
0347-241       &  3:49:15.39 & -24:01:12.58 & 1.88 & 17.5 &  159.0 &  290.2 &   21.54 &   29.25 &  240.81 &  259.02 \\ 
0352+123       &  3:55:45.60 &  12:31:45.70 & 1.61 & 19.3 &  316.0 &  608.4 &  112.22 &  190.84 &  315.87 &  554.11 \\ 
0414+0534      &  4:14:37.74 &   5:34:43.75 & 2.64 &  \nodata &  959.0 & 2676.0 &  188.37 &  466.17 & 1440.80 & 1888.40 \\ 
0421+019       &  4:24: 8.56 &   2:04:24.90 & 2.06 & 17.0 &  711.0 & 1150.2 &  640.91 &  642.77 & 1192.30 & 1199.00 \\ 
0424-131       &  4:27: 7.34 & -13:02:53.60 & 2.17 & 17.5 &  280.0 & 1087.0 &  123.98 &  129.07 & 1088.10 & 1088.90 \\ 
0445+097       &  4:48:21.78 &   9:50:50.00 & 2.11 & 19.6 &  543.0 & 1007.0 &  269.75 &  412.26 &    4.93 &  193.95 \\ 
0446-208       &  4:48:58.85 & -20:44:45.60 & 1.90 & 17.0 &  100.0 &  250.2 &   50.66 &   52.35 &   98.62 &  250.62 \\ 
0457+024       &  4:59:52.06 &   2:29:31.50 & 2.38 & 19.4 & 1782.0 & 2067.3 &  870.98 &  868.81 &  849.07 &  912.86 \\ 
0458-020       &  5:01:12.75 &  -1:59:14.90 & 2.29 & 18.4 & 1650.0 & 2373.5 & 1090.00 & 1139.20 & 1314.80 & 1481.20 \\ 
0528-250       &  5:30: 7.93 & -25:03:30.10 & 2.78 & 17.2 &  845.0 & 1162.3 &  666.89 &  665.62 & 1053.30 & 1053.00 \\ 
0636+680       &  6:42: 4.01 &  67:58:35.10 & 3.17 & 16.5 &  488.0 &  193.0 &  342.44 &  342.52 &  202.86 &  202.12 \\ 
0642+449       &  6:46:32.02 &  44:51:16.90 & 3.41 & 18.4 & 1220.0 &  452.9 & 3228.60 & 3225.30 &  471.30 &  471.03 \\ 
0731+653       &  7:36:21.03 &  65:13:10.50 & 3.04 & 18.5 &   72.0 &   85.0 &   64.23 &   65.58 &   66.25 &   79.65 \\ 
0736-063       &  7:38:57.13 &  -6:26:57.30 & 1.90 & 18.5 & 1190.0 & 1600.9 &  609.37 &  634.20 & 1594.30 & 1623.70 \\ 
0747+613       &  7:52:11.99 &  61:12:23.30 & 2.49 & 17.5 &  307.0 &  875.8 &  138.20 &  145.18 &  856.10 &  874.34 \\ 
0802+103       &  8:04:47.96 &  10:15:22.40 & 1.96 & 18.2 &  539.0 & 1849.9 &    \nodata &    \nodata &    \nodata & 1879.20 \\ 
0805+046       &  8:07:57.51 &   4:32:34.60 & 2.88 & 18.4 &  350.0 &  505.0 &  328.75 &  317.73 &  359.76 &  547.02 \\ 
0830+115       &  8:33:14.38 &  11:23:36.50 & 2.98 & 18.5 &  240.0 &  428.9 &  272.00 &  269.77 &  421.35 &  426.10 \\ 
0835+580       &  8:39: 6.48 &  57:54:17.50 & 1.54 & 17.6 &  649.0 & 2258.6 &   12.70 &   55.27 &   34.85 & 2267.90 \\ 
0836+195       &  8:39: 6.97 &  19:21:48.40 & 1.69 & 17.6 &  159.0 &  149.2 &    \nodata &    \nodata &   85.47 &  461.83 \\ 
0843+136       &  8:45:47.25 &  13:28:58.70 & 1.88 & 17.8 &  200.0 &  472.0 &    3.29 &    3.43 &  344.75 &  440.21 \\ 
0938+119       &  9:41:13.56 &  11:45:32.20 & 3.19 & 19.0 &  123.0 &  291.0 &    \nodata &    \nodata &  248.62 &  258.43 \\ 
0941+261       &  9:44:42.32 &  25:54:43.20 & 2.91 & 18.7 &  306.0 &  762.0 &   75.15 &   86.10 &  699.93 &  797.24 \\ 
0959+105       & 10:01:57.71 &  10:15:49.90 & 1.53 & 17.5 &  301.0 &  283.0 &  289.02 &  284.80 &  323.63 &  323.56 \\ 
1011+250       & 10:13:53.45 &  24:49:17.30 & 1.63 & 15.4 &  937.0 &  483.3 &    \nodata &    \nodata &  384.85 &  419.14 \\ 
1017+1055      & 10:20:10.04 &  10:40: 2.20 & 3.16 & 18.3 &  191.0 &  538.0 &    \nodata &    \nodata &  208.54 &  543.27 \\ 
1033+1342      & 10:36:26.90 &  13:26:52.00 & 3.09 & 18.0 &   93.0 &  252.0 &    \nodata &    \nodata &   57.85 &  112.23 \\ 
1055+499       & 10:58:13.02 &  49:39:35.80 & 2.40 & 19.5 &  100.0 &  246.9 &   52.24 &   59.75 &  182.87 &  236.67 \\ 
1100-264       & 11:03:25.30 & -26:45:15.40 & 2.15 & 16.0 &    \nodata &   15.5 &    \nodata &    \nodata &   10.12 &   12.53 \\ 
1116+128       & 11:18:57.32 &  12:34:41.40 & 2.12 & 18.5 & 1820.0 & 1990.0 & 1184.40 & 1269.60 & 2030.00 & 2283.20 \\ 
1124+571       & 11:27:40.13 &  56:50:15.10 & 2.89 & 19.0 &  448.0 &  475.7 &  187.80 &  190.91 &  470.28 &  479.05 \\ 
1126+101       & 11:29:14.10 &   9:51:59.80 & 1.52 & 18.0 &  367.0 &  552.0 &   65.47 &   68.91 &  212.94 &  589.81 \\ 
1148-001       & 11:50:43.89 &   0:23:55.20 & 1.98 & 17.3 & 2038.0 & 2684.0 &  259.84 &  251.66 & 2758.40 & 2835.70 \\ 
1157+014       & 11:59:44.76 &   1:12: 7.40 & 1.99 & 17.7 &  111.0 &  270.0 &   47.08 &   42.14 &  274.18 &  268.20 \\ 
1211+334       & 12:14: 4.16 &  33:09:45.80 & 1.59 & 17.9 &  627.0 & 1404.2 &  508.20 &  548.00 & 1383.90 & 1416.60 \\ 
1214+106       & 12:17: 1.34 &  10:19:53.70 & 1.88 & 18.5 &   59.0 &  144.0 &    2.90 &    5.62 &    \nodata &    \nodata \\ 
1215+333       & 12:17:32.60 &  33:05:38.30 & 2.61 & 17.5 &  117.0 &  193.8 &   11.87 &   12.49 &  177.31 &  185.62 \\ 
1218+3359      & 12:20:33.88 &  33:43:11.80 & 1.52 & 18.6 &  842.0 & 2846.5 &  102.88 &  445.09 &  141.40 & 2834.40 \\ 
1221+113       & 12:24:19.86 &  11:07:22.80 & 1.76 & 18.7 &  132.0 &  354.0 &   20.80 &   24.22 &  304.79 &  338.91 \\ 
1222+228       & 12:25:27.40 &  22:35:12.70 & 2.05 & 15.5 &   12.0 &    6.7 &    1.95 &    2.15 &    3.77 &   12.38 \\ 
1225+317       & 12:28:24.84 &  31:28:37.80 & 2.22 & 15.9 &  345.0 &  334.0 &  365.89 &  367.51 &  329.64 &  345.28 \\ 
1226+105       & 12:28:36.91 &  10:18:41.60 & 2.31 & 18.5 &  226.0 &  657.0 &   14.03 &  116.07 &   91.81 &  651.68 \\ 
1256+357       & 12:58:29.77 &  35:28:43.00 & 1.89 & 18.2 &    6.0 &   19.2 &    5.15 &    4.27 &   12.83 &   23.43 \\ 
1302-1404      & 13:05:25.24 & -14:20:41.40 & 4.00 & 18.6 &    \nodata &   20.7 &    2.59 &    2.99 &   20.40 &   20.49 \\ 
1308+182       & 13:10:56.69 &  17:59:37.90 & 1.68 & 17.5 &  154.0 &  360.0 &   30.19 &   96.54 &   26.36 &  444.28 \\ 
1311-270       & 13:13:47.32 & -27:16:48.50 & 2.20 & 17.4 &  249.0 &  635.6 &    \nodata &    \nodata &   41.09 &  567.60 \\ 
1313+200       & 13:16:24.61 &  19:47: 4.00 & 2.46 & 18.5 &  332.0 &  420.0 &  236.87 &  254.23 &  312.04 &  344.24 \\ 
1318+113       & 13:21:18.86 &  11:06:50.10 & 2.17 & 19.1 &  918.0 & 2251.0 &   53.54 &  192.51 & 1196.90 & 2258.00 \\ 
1323+655       & 13:25:29.63 &  65:15:14.50 & 1.62 & 17.8 &  225.0 &  713.9 &    7.99 &  112.19 &    \nodata &  695.92 \\ 
1331+170       & 13:33:35.89 &  16:49: 3.40 & 2.08 & 16.7 &  708.0 &  292.0 &  480.07 &  480.74 &  381.54 &  408.96 \\ 
1343+2640A     & 13:45:43.77 &  26:25: 6.60 & 2.03 & 20.2 &    \nodata &    8.9 &    \nodata &    \nodata &    6.67 &   19.25 \\ 
1354+258       & 13:57: 6.55 &  25:37:29.20 & 2.01 & 18.5 &  114.0 &  398.5 &    \nodata &    \nodata &  124.97 &  391.09 \\ 
1402-012       & 14:04:45.91 &  -1:30:22.00 & 2.52 & 17.2 &  810.0 &  620.0 &  318.25 &  319.80 &  486.21 &  481.34 \\ 
1421+122       & 14:23:30.13 &  11:59:51.80 & 1.61 & 18.0 &  678.0 &  913.0 &  251.20 &  295.30 &  816.98 &  974.45 \\ 
1421+330       & 14:23:26.07 &  32:52:20.80 & 1.90 & 16.7 &    9.0 &    8.6 &   12.30 &   12.76 &    8.57 &   10.51 \\ 
1435+638       & 14:36:45.72 &  63:36:38.00 & 2.07 & 15.0 &  757.0 &  951.8 &   72.21 &  141.77 &  881.06 &  871.52 \\ 
1442+101       & 14:45:16.47 &   9:58:36.20 & 3.54 & 17.3 & 1275.0 & 2415.0 &  631.72 &  631.22 & 2518.40 & 2516.60 \\ 
1448-232       & 14:51: 2.49 & -23:29:31.80 & 2.21 & 17.0 &  284.0 &  439.5 &  141.83 &  141.58 &  475.28 &  495.29 \\ 
1508+5714      & 15:10: 2.91 &  57:02:43.20 & 4.28 & 18.9 &  292.0 &  202.5 &    \nodata &    \nodata &  286.45 &  295.01 \\ 
1511+103       & 15:13:29.33 &  10:11: 5.70 & 1.55 & 17.7 &   67.0 &  185.0 &   10.42 &   11.18 &   20.19 &  203.13 \\ 
1542+042       & 15:44:59.51 &   4:07:46.00 & 2.18 & 18.0 &  387.0 &  766.0 &   24.03 &  294.24 &  761.41 &  820.84 \\ 
1548+114B      & 15:50:43.94 &  11:20:46.50 & 1.90 & 19.0 &  543.0 &  598.0 &    \nodata &    \nodata &    \nodata &  393.43 \\ 
1556+335       & 15:58:55.18 &  33:23:18.80 & 1.65 & 17.0 &   95.0 &  142.7 &  119.26 &  121.81 &  143.84 &  144.30 \\ 
1559+173       & 16:01:20.38 &  17:14:16.40 & 1.94 & 17.7 &  294.0 &  824.0 &  154.31 &  184.21 &  714.62 &  798.95 \\ 
1602-001       & 16:04:56.15 &   0:19: 7.30 & 1.62 & 17.5 &  343.0 & 1007.0 &   11.52 &   12.48 &  326.65 & 1021.40 \\ 
1602+576       & 16:03:55.85 &  57:30:54.10 & 2.86 & 18.3 &  356.0 &  787.0 &   34.70 &   87.80 &  361.32 &  363.74 \\ 
1606+289       & 16:08:11.26 &  28:49: 2.20 & 1.98 & 19.0 &  145.0 &  600.4 &    1.33 &   66.17 &    1.73 &  611.52 \\ 
1607+183       & 16:10: 5.28 &  18:11:45.40 & 3.12 & 18.5 &  131.0 &  248.0 &   48.47 &   61.30 &  175.34 &  215.97 \\ 
1614+051       & 16:16:37.56 &   4:59:33.30 & 3.22 & 19.5 &  916.0 &  306.0 &  107.04 &  198.83 &  348.59 &  373.95 \\ 
1629+120       & 16:31:45.22 &  11:56: 3.20 & 1.79 & 18.5 &  864.0 & 1628.0 &  153.76 &  427.52 & 1412.50 & 1719.40 \\ 
1629+680       & 16:29:51.78 &  67:57:15.20 & 2.48 & 18.7 &  409.0 &  849.1 &   25.08 &   42.09 &  904.77 &  909.28 \\ 
1701+379       & 17:03: 8.05 &  37:51:26.40 & 2.46 & 19.0 &   33.0 &  111.1 &    0.75 &   20.83 &    6.88 &  119.35 \\ 
1702+289       & 17:04: 7.18 &  29:46:59.60 & 1.93 & 19.1 &  616.0 & 1414.5 &    \nodata &  197.37 &  896.55 & 1421.30 \\ 
1705+018       & 17:07:34.41 &   1:48:46.80 & 2.58 & 18.9 &  436.0 &  405.0 &    \nodata &    \nodata &  483.92 &  493.40 \\ 
1718+4807      & 17:19:38.25 &  48:04:12.70 & 1.08 & 15.3 &  109.0 &   63.2 &  212.87 &  213.71 &   58.96 &   62.21 \\ 
1726+344       & 17:27:49.83 &  34:22:40.00 & 2.43 & 18.5 &   23.0 &   87.6 &    \nodata &    \nodata &   74.61 &   80.08 \\ 
1738+350       & 17:40:20.24 &  35:00:47.70 & 3.24 & 20.5 &   57.0 &  132.0 &   22.29 &   22.00 &   85.70 &  139.41 \\ 
1756+237       & 17:59: 0.93 &  23:43:46.20 & 1.72 & 18.0 &  929.0 &  604.0 &   73.69 &  238.52 &  602.22 &  636.84 \\ 
1816+475       & 18:18:19.56 &  47:36:45.10 & 2.23 & 18.2 &  150.0 &  545.0 &    \nodata &    \nodata &  186.65 &  574.05 \\ 
2044-168       & 20:47:19.63 & -16:39: 5.60 & 1.94 & 17.4 &  652.0 &  745.4 &  319.27 &  327.06 &  530.03 &  766.19 \\ 
2048+196       & 20:51:12.82 &  19:50: 6.20 & 2.37 & 18.5 &  105.0 &  149.0 &   39.48 &   37.75 &  177.57 &  179.25 \\ 
2120+168       & 21:22:46.32 &  17:04:38.40 & 1.80 & 18.0 &  394.0 & 1672.0 &    2.06 &  184.25 &    8.29 & 1560.90 \\ 
2121+0522      & 21:23:44.52 &   5:35:21.60 & 1.94 & 17.5 & 2784.0 & 1142.0 & 2162.20 & 2165.80 & 1210.50 & 1227.30 \\ 
2126-158       & 21:29:12.09 & -15:38:42.30 & 3.27 & 17.3 & 1186.0 &  590.2 & 1127.90 & 1132.90 &  476.38 &  446.45 \\ 
2136+141       & 21:39: 1.34 &  14:23:35.90 & 2.43 & 18.5 & 1073.0 & 1150.0 & 2135.10 & 2132.10 & 1140.50 & 1133.60 \\ 
2146-133       & 21:49:28.68 & -13:04:25.90 & 1.80 & 19.5 &  670.0 & 1820.7 &   26.30 &  285.43 &  892.61 & 1719.20 \\ 
2149+212       & 21:51:45.90 &  21:30:13.40 & 1.54 & 19.0 &  377.0 &  987.2 &   88.30 &  183.27 &  860.59 &  969.74 \\ 
2150+053       & 21:53:24.51 &   5:36:18.60 & 1.98 & 17.8 &  427.0 & 1292.0 &   14.24 &  158.68 &   35.80 & 1146.70 \\ 
2156+297       & 21:58:41.97 &  29:59: 9.40 & 1.76 & 17.5 &  413.0 & 1223.0 &  112.37 &  229.31 &  885.34 & 1187.90 \\ 
2158+101       & 22:01:16.67 &  10:23:47.40 & 1.73 & 17.7 &  212.0 &  420.0 &   45.46 &   97.41 &  497.62 &  558.36 \\ 
2222+051       & 22:25:14.75 &   5:27: 8.80 & 2.33 & 18.5 &  266.0 &  843.0 &   45.18 &  124.28 &  548.42 &  854.74 \\ 
2223+210       & 22:25:38.05 &  21:18: 6.10 & 1.95 & 18.2 & 1024.0 & 1837.7 &  767.86 &  820.06 & 1829.80 & 1976.40 \\ 
2248+192       & 22:50:32.83 &  19:31:19.80 & 1.80 & 18.5 &  250.0 &  871.0 &    5.22 &  147.23 &  452.40 &  861.95 \\ 
2249+185       & 22:51:34.76 &  18:48:40.00 & 1.76 & 18.5 &  797.0 & 2133.5 &  345.18 &  469.33 &  430.57 &  458.18 \\ 
2251+244       & 22:54: 9.27 &  24:45:24.10 & 2.33 & 17.8 &  869.0 & 2064.0 &  664.50 &  717.59 & 1788.20 & 1802.30 \\ 
2254+0257      & 22:57:17.57 &   2:43:17.10 & 2.09 & 17.1 &  503.0 &  236.0 &  430.03 &  431.95 &  172.06 &  171.64 \\ 
2338+042       & 23:40:57.98 &   4:31:15.30 & 2.59 & 19.5 &  453.0 & 1521.0 &  103.62 &  231.52 & 1140.50 & 1552.30 \\ 
2345+061       & 23:48:31.77 &   6:24:59.20 & 1.54 & 17.5 &  343.0 &  655.0 &   82.92 &  183.79 &  580.07 &  660.67 \\ 
2351-154       & 23:54:30.17 & -15:13:11.20 & 2.67 & 17.0 &  970.0 &  865.2 &  812.46 &  819.31 &  949.50 &  973.05 \\ 
2354+144       & 23:57:18.32 &  14:46: 9.00 & 1.82 & 18.1 &  458.0 & 1094.0 &   21.81 &  108.72 &  101.35 &  969.82 \\ 
\enddata
\end{deluxetable}

\begin{deluxetable}{lrrrrrrrrr}
\tabletypesize{\small}
\tablewidth{0pt}
\tablecaption{Derived Quasar Properties \label{tab:tab2}}
\tablehead{
\colhead{Name} & 
\colhead{$\alpha_6^{20}$} &
\colhead{$\alpha_{3.5}^{20}$} &
\colhead{$\sigma_{\alpha_{3.5}^{20}}$} &
\colhead{$\alpha_{3.5}^{20}$} &
\colhead{${\rm R}_V$} &
\colhead{${\rm C}_{20}$} &
\colhead{${\rm L}_{20}$} &
\colhead{${\rm L}_{20}$} &
\colhead{${\rm M}_V$} \\
\colhead{} & 
\colhead{} &
\colhead{(core)} &
\colhead{(core)} &
\colhead{(total)} &
\colhead{} &
\colhead{} &
\colhead{(core)} &
\colhead{(total)} &
\colhead{}
}
\startdata
FBQS0047-0156 & \nodata & \nodata & \nodata & \nodata & \nodata &   \nodata &   \nodata &   32.67 &  -27.56 \\ 
FBQS0048-0103  & \nodata & \nodata & \nodata & \nodata & \nodata &   \nodata &   \nodata &   32.47 &  -27.19 \\ 
FBQS0210-0152 & \nodata & \nodata & \nodata & \nodata & \nodata &   \nodata &   \nodata &   32.31 &  -27.10 \\ 
FBQS0256-0119 & \nodata & -0.69 & 0.12 & -0.69 &  1.92 &   \nodata &   33.64 &   33.46 &  -27.30 \\ 
FBQS0729+2524 &  1.22 &  0.28 & 0.10 & -0.55 &  1.21 &   \nodata &   33.05 &   32.19 &  -27.62 \\ 
FBQS0804+2516 & \nodata & -1.05 & 0.51 & -1.18 &  1.08 &   \nodata &   32.86 &   32.72 &  -27.47 \\
FBQS0857+3313 &  0.27 &  0.32 & 0.11 &  0.43 &  1.01 &   \nodata &   32.95 &   33.03 &  -27.86 \\ 
FBQS0934+3153  & \nodata &  0.02 & 0.22 & -0.31 &  0.66 &    0.54 &   32.47 &   32.89 &  -27.54 \\ 
FBQS0955+333  &  0.39 &  0.57 & 0.09 &  0.49 &  1.01 &   \nodata &   33.20 &   33.20 &  -28.46 \\ 
FBQS1045+3440 & -0.09 & -0.43 & 0.11 & -0.35 &  1.61 &    2.73 &   33.37 &   33.22 &  -27.41 \\ 
FBQS1253+2905  & -0.42 & -0.22 & 0.10 & -0.27 &  1.79 &    2.14 &   33.82 &   33.97 &  -28.05 \\ 
FBQS1348+2840  &  0.16 & -0.27 & 0.10 & -0.21 &  1.87 &   \nodata &   33.81 &   33.73 &  -27.83 \\ 
FBQS1457+2707  &  1.99 & \nodata & \nodata & \nodata & \nodata &   \nodata &   \nodata &   31.24 &  -28.06 \\ 
FBQS1537+2716 & \nodata &  0.27 & 0.14 &  0.09 &  0.06 &   \nodata &   32.14 &   32.28 &  -28.20 \\ 
FBQS1540+4138 &  0.00 &  0.40 & 0.09 &  0.45 &  0.85 &   \nodata &   32.99 &   33.19 &  -28.35 \\ 
FBQS1625+2646 & \nodata & \nodata & \nodata & \nodata & \nodata &   \nodata &   \nodata &   32.92 &  -28.27 \\ 
FBQS1634+3203 &  0.27 &  0.19 & 0.09 &  0.18 &  1.85 &    2.12 &   33.92 &   33.95 &  -28.19 \\ 
FBQS1651+4002 &  0.10 & -1.29 & 0.16 & -0.92 &  2.09 &   \nodata &   34.19 &   33.43 &  -28.25 \\ 
FBQS2216-0057  & \nodata &  0.45 & 0.12 &  0.39 &  0.66 &    3.00 &   32.42 &   32.94 &  -27.42 \\ 
FBQS2233-0838 & \nodata & -0.41 & 0.14 & -0.44 &  1.37 &   43.06 &   33.03 &   33.02 &  -27.14 \\ 
0002+051       &  0.68 &  0.12 & 0.09 &  0.11 &  1.57 &   \nodata &   33.81 &   33.49 &  -28.66 \\ 
0004+171       & -0.28 & -0.32 & 0.09 & -0.41 &  2.68 &    0.55 &   34.41 &   34.51 &  -27.30 \\ 
0014+813       & -0.18 & -0.20 & 0.09 & -0.19 &  2.37 &   \nodata &   35.06 &   35.04 &  -29.64 \\ 
0017+154       & -1.21 & \nodata & \nodata & -1.37 & \nodata &   \nodata &   \nodata &   35.66 &  -26.79 \\ 
0033+0951      & -0.63 &  0.17 & 0.09 & -0.30 &  2.28 &    0.21 &   34.01 &   34.85 &  -27.38 \\ 
0038-019       & -0.73 &  0.00 & 0.12 & -1.08 &  2.00 &    0.01 &   33.20 &   34.94 &  -26.08 \\ 
0051+291       & -0.56 & \nodata & \nodata & -0.57 & \nodata &   \nodata &   \nodata &   34.64 &  -26.98 \\ 
0109+176       & -0.90 & -0.06 & 0.10 & -0.88 &  1.77 &    0.03 &   33.42 &   34.93 &  -27.15 \\ 
0119-046       & -0.51 & -0.60 & 0.09 & -0.67 &  2.84 &    0.36 &   35.01 &   35.14 &  -28.48 \\ 
0123+257       &  0.21 & -0.24 & 0.09 & -0.26 &  3.01 &    2.89 &   34.94 &   34.74 &  -27.85 \\ 
0136+1737      & -0.29 & -0.14 & 0.09 & -0.14 &  2.89 &    0.91 &   34.56 &   34.74 &  -27.16 \\ 
0149+336       & -0.15 &  0.02 & 0.09 & -0.25 &  2.81 &    0.24 &   34.38 &   34.82 &  -26.91 \\ 
0150-202       &  1.23 &  0.84 & 0.10 &  0.84 &  0.80 &   \nodata &   32.80 &   32.59 &  -28.04 \\ 
0201+3634      & -0.37 & -0.50 & 0.09 & -0.50 &  2.84 &    0.90 &   34.98 &   34.96 &  -28.31 \\ 
0225-014       & -0.55 & \nodata & \nodata & -0.89 &  2.47 &    0.07 &   34.00 &   34.67 &  -26.87 \\ 
0226-038       & -0.40 &  0.03 & 0.09 & -0.14 &  2.45 &    0.44 &   34.47 &   34.92 &  -28.09 \\ 
0229+1309      &  0.36 & -0.06 & 0.09 & -0.08 &  3.22 &    3.27 &   34.94 &   34.78 &  -27.35 \\ 
0237-233       & -0.55 & -1.47 & 0.10 & -1.30 &  4.11 &    1.27 &   36.33 &   35.89 &  -28.59 \\ 
0238+100       & -0.92 & -0.36 & 0.10 & -0.94 &  1.95 &    0.04 &   33.44 &   34.53 &  -26.78 \\ 
0256-000       &  0.18 &  1.04 & 0.24 &  0.84 & -0.06 &    1.99 &   31.73 &   32.47 &  -27.42 \\ 
0316-203       & -0.74 & -1.44 & 0.13 & -1.58 &  3.42 &    0.18 &   34.75 &   34.42 &  -26.28 \\ 
0329-2534      & -0.19 & -0.85 & 0.17 & -1.08 &  2.75 &    0.29 &   34.81 &   34.55 &  -28.13 \\ 
0336-0142      & -0.18 & -0.26 & 0.09 & -0.35 &  3.30 &    0.68 &   34.89 &   34.93 &  -26.92 \\ 
0347-241       & -0.48 & -1.36 & 0.13 & -1.22 &  2.99 &    0.40 &   34.70 &   34.38 &  -27.35 \\ 
0352+123       & -0.53 & \nodata & \nodata & -0.60 &  3.41 &    0.28 &   34.26 &   34.59 &  -25.20 \\ 
0414+0534      & -0.83 & \nodata & \nodata & -0.79 & \nodata &   \nodata &   \nodata &   35.82 &   \nodata \\ 
0421+019       & -0.39 & -0.35 & 0.09 & -0.35 &  3.02 &   \nodata &   35.00 &   35.01 &  -28.00 \\ 
0424-131       & -1.09 & -1.22 & 0.09 & -1.20 &  3.59 &   \nodata &   35.44 &   35.37 &  -27.66 \\ 
0445+097       & -0.50 & \nodata & \nodata &  0.42 & \nodata &   \nodata &   \nodata &   35.02 &  -25.55 \\ 
0446-208       & -0.74 & -0.37 & 0.09 & -0.88 &  1.94 &    0.15 &   33.87 &   34.44 &  -27.86 \\ 
0457+024       & -0.12 &  0.01 & 0.09 & -0.03 &  3.59 &   \nodata &   34.78 &   35.23 &  -25.97 \\ 
0458-020       & -0.29 & -0.11 & 0.11 & -0.15 &  3.46 &   \nodata &   35.00 &   35.35 &  -26.88 \\ 
0528-250       & -0.26 & -0.26 & 0.09 & -0.26 &  2.93 &    1.82 &   35.13 &   35.17 &  -28.47 \\ 
0636+680       &  0.75 &  0.29 & 0.09 &  0.30 &  1.53 &   \nodata &   34.17 &   33.86 &  -29.54 \\ 
0642+449       &  0.80 &  1.08 & 0.09 &  1.08 &  2.14 &   \nodata &   34.07 &   34.23 &  -27.74 \\ 
0731+653       & -0.13 & -0.02 & 0.09 & -0.11 &  2.07 &    0.85 &   33.85 &   34.02 &  -27.41 \\ 
0736-063       & -0.24 & -0.54 & 0.11 & -0.53 &  3.83 &   46.85 &   35.15 &   35.02 &  -26.37 \\ 
0747+613       & -0.84 & -1.02 & 0.09 & -1.01 &  3.39 &    3.46 &   35.37 &   35.28 &  -27.97 \\ 
0802+103       & -0.99 & \nodata & \nodata & \nodata & \nodata &   \nodata &   \nodata &   35.45 &  -26.74 \\ 
0805+046       & -0.30 & -0.05 & 0.09 & -0.31 &  2.78 &    0.60 &   34.57 &   34.86 &  -27.44 \\ 
0830+115       & -0.47 & -0.25 & 0.09 & -0.26 &  3.01 &   10.00 &   34.77 &   34.92 &  -27.36 \\ 
0835+580       & -1.00 & -0.57 & 0.12 & -2.09 &  1.86 &    0.00 &   33.32 &   35.31 &  -26.78 \\ 
0836+195       &  0.05 & \nodata & \nodata & \nodata &  2.15 &   \nodata &   33.72 &   33.77 &  -27.01 \\ 
0843+136       & -0.70 & -2.61 & 0.40 & -2.72 &  3.84 &    0.06 &   35.43 &   34.69 &  -27.04 \\ 
0938+119       & -0.69 & \nodata & \nodata & \nodata &  3.06 &    0.79 &   34.69 &   34.94 &  -27.01 \\ 
0941+261       & -0.73 & -1.25 & 0.09 & -1.25 &  3.91 &    0.52 &   35.58 &   35.30 &  -27.11 \\ 
0959+105       &  0.05 & -0.06 & 0.09 & -0.07 &  2.57 &   \nodata &   34.09 &   33.98 &  -26.89 \\ 
1011+250       &  0.53 & \nodata & \nodata & \nodata &  1.93 &    1.01 &   34.35 &   34.06 &  -29.13 \\ 
1017+1055      & -0.80 & \nodata & \nodata & \nodata & \nodata &   \nodata &   \nodata &   35.27 &  -27.69 \\ 
1033+1342      & -0.80 & \nodata & \nodata & \nodata &  2.04 &   \nodata &   34.03 &   34.92 &  -27.94 \\ 
1055+499       & -0.73 & -0.70 & 0.09 & -0.77 &  3.34 &    0.36 &   34.49 &   34.64 &  -25.88 \\ 
1100-264       & \nodata & \nodata & \nodata & \nodata &  0.55 &    0.38 &   32.99 &   33.18 &  -29.12 \\ 
1116+128       & -0.07 & -0.30 & 0.11 & -0.33 &  3.79 &   \nodata &   35.23 &   35.11 &  -26.64 \\ 
1124+571       & -0.05 & -0.52 & 0.09 & -0.52 &  3.43 &   11.08 &   34.96 &   34.69 &  -26.80 \\ 
1126+101       & -0.33 & -0.66 & 0.09 & -1.21 &  2.83 &    0.14 &   34.14 &   34.42 &  -26.37 \\ 
1148-001       & -0.20 & -1.33 & 0.12 & -1.36 &  3.97 &   \nodata &   35.80 &   35.25 &  -27.62 \\ 
1157+014       & -0.72 & -0.99 & 0.09 & -1.04 &  2.97 &   \nodata &   34.64 &   34.51 &  -27.23 \\ 
1211+334       & -0.65 & -0.56 & 0.09 & -0.53 &  3.56 &   15.38 &   34.95 &   35.00 &  -26.59 \\ 
1214+106       & -0.70 & \nodata & \nodata & \nodata & \nodata &   \nodata &   \nodata &   34.17 &  -26.35 \\ 
1215+333       & -0.41 & -1.52 & 0.11 & -1.52 &  2.98 &    0.43 &   35.01 &   34.43 &  -28.07 \\ 
1218+3359      & -0.99 & -0.18 & 0.12 & -1.04 &  2.70 &    0.02 &   33.77 &   35.40 &  -25.77 \\ 
1221+113       & -0.79 & -1.51 & 0.10 & -1.48 &  3.63 &    0.48 &   34.80 &   34.55 &  -26.00 \\ 
1222+228       &  0.47 & -0.37 & 0.25 & -0.98 & -0.09 &   \nodata &   32.51 &   32.35 &  -29.54 \\ 
1225+317       &  0.03 &  0.06 & 0.09 &  0.04 &  1.77 &   25.15 &   34.30 &   34.32 &  -29.34 \\ 
1226+105       & -0.86 & \nodata & \nodata & -0.97 & \nodata &   \nodata &   \nodata &   35.10 &  -26.80 \\ 
1256+357       & -0.94 & -0.51 & 0.13 & -0.96 &  1.62 &    0.40 &   33.04 &   33.42 &  -26.62 \\ 
1302-1404      & \nodata & -1.16 & 0.14 & -1.08 &  2.30 &    2.13 &   34.31 &   33.79 &  -27.91 \\ 
1308+182       & -0.68 &  0.08 & 0.10 & -0.86 &  1.40 &    0.03 &   33.00 &   34.46 &  -27.09 \\ 
1311-270       & -0.75 & \nodata & \nodata & \nodata &  1.72 &    0.01 &   33.62 &   34.98 &  -27.76 \\ 
1313+200       & -0.20 & -0.15 & 0.09 & -0.17 &  2.88 &    0.69 &   34.45 &   34.61 &  -26.94 \\ 
1318+113       & -0.72 & \nodata & \nodata & -1.38 & \nodata &   \nodata &   \nodata &   35.51 &  -26.04 \\ 
1323+655       & -0.93 & \nodata & \nodata & -1.02 & \nodata &   \nodata &   \nodata &   34.83 &  -26.72 \\ 
1331+170       &  0.70 &  0.13 & 0.09 &  0.09 &  2.15 &   \nodata &   34.28 &   33.89 &  -28.36 \\ 
1343+2640A     &  0.00 & \nodata & \nodata & \nodata &  2.07 &   \nodata &   32.76 &   32.70 &  -24.78 \\ 
1354+258       & -1.01 & \nodata & \nodata & \nodata & \nodata &   \nodata &   \nodata &   34.82 &  -26.49 \\ 
1402-012       &  0.22 & -0.24 & 0.09 & -0.23 &  2.57 &   \nodata &   34.71 &   34.57 &  -28.34 \\ 
1421+122       & -0.24 & -0.66 & 0.09 & -0.67 &  3.43 &    1.73 &   34.78 &   34.65 &  -26.46 \\ 
1421+330       &  0.04 &  0.20 & 0.12 &  0.11 &  0.50 &  107.13 &   32.54 &   32.62 &  -28.17 \\ 
1435+638       & -0.18 & -1.40 & 0.14 & -1.02 &  2.58 &    0.84 &   35.39 &   34.83 &  -30.06 \\ 
1442+101       & -0.51 & -0.78 & 0.09 & -0.78 &  3.60 &   \nodata &   36.02 &   35.83 &  -28.98 \\ 
1448-232       & -0.35 & -0.68 & 0.09 & -0.70 &  2.74 &   \nodata &   34.83 &   34.63 &  -28.25 \\ 
1508+5714      &  0.29 & \nodata & \nodata & \nodata &  3.02 &   \nodata &   34.98 &   34.33 &  -27.76 \\ 
1511+103       & -0.80 & -0.37 & 0.11 & -1.63 &  1.58 &    0.03 &   33.01 &   34.15 &  -26.68 \\ 
1542+042       & -0.55 & -1.94 & 0.10 & -0.58 &  3.99 &    5.52 &   35.65 &   34.96 &  -27.18 \\ 
1548+114B      & -0.08 & \nodata & \nodata & \nodata & \nodata &   \nodata &   \nodata &   34.52 &  -25.87 \\ 
1556+335       & -0.33 & -0.11 & 0.09 & -0.10 &  2.02 &   \nodata &   33.81 &   33.90 &  -27.56 \\ 
1559+173       & -0.83 & \nodata & \nodata & -0.82 & \nodata &   \nodata &   \nodata &   35.02 &  -27.21 \\ 
1602-001       & -0.90 & -1.88 & 0.20 & -2.47 &  3.32 &    0.03 &   34.89 &   34.97 &  -27.03 \\ 
1602+576       & -0.60 & -1.32 & 0.11 & -0.80 &  3.50 &   \nodata &   35.31 &   35.22 &  -27.47 \\ 
1606+289       & -1.14 & -0.15 & 0.70 & -1.25 &  0.88 &    0.00 &   32.04 &   35.05 &  -25.96 \\ 
1607+183       & -0.51 & -0.72 & 0.09 & -0.71 &  2.91 &    0.21 &   34.72 &   34.74 &  -27.47 \\ 
1614+051       &  0.88 & -0.66 & 0.11 & -0.35 &  3.57 &   \nodata &   35.01 &   33.99 &  -26.53 \\ 
1629+120       & -0.51 & \nodata & \nodata & -0.78 &  3.72 &    1.56 &   34.99 &   35.10 &  -26.24 \\ 
1629+680       & -0.59 & -2.01 & 0.12 & -1.73 &  4.42 &   \nodata &   35.93 &   35.13 &  -26.76 \\ 
1701+379       & -0.98 & -1.25 & 0.53 & -0.98 &  2.01 &    0.00 &   33.39 &   34.45 &  -26.44 \\ 
1702+289       & -0.67 & \nodata & \nodata & -1.11 &  3.77 &    0.38 &   34.85 &   35.17 &  -25.76 \\ 
1705+018       &  0.06 & \nodata & \nodata & \nodata &  3.35 &   \nodata &   34.81 &   34.48 &  -26.64 \\ 
1718+4807      &  0.44 &  0.72 & 0.09 &  0.69 &  0.79 &   11.34 &   32.84 &   32.96 &  -28.29 \\ 
1726+344       & -1.08 & \nodata & \nodata & \nodata &  2.39 &    1.02 &   33.96 &   34.39 &  -26.91 \\ 
1738+350       & -0.68 & -0.76 & 0.10 & -1.04 &  3.42 &    0.15 &   34.46 &   34.60 &  -25.55 \\ 
1756+237       &  0.35 & \nodata & \nodata & -0.55 &  3.16 &   83.31 &   34.59 &   34.26 &  -26.65 \\ 
1816+475       & -1.04 & \nodata & \nodata & \nodata & \nodata &   \nodata &   \nodata &   35.08 &  -27.02 \\ 
2044-168       & -0.11 & -0.28 & 0.11 & -0.48 &  2.77 &    0.62 &   34.57 &   34.64 &  -27.55 \\ 
2048+196       & -0.30 & -0.84 & 0.09 & -0.87 &  3.01 &   \nodata &   34.55 &   34.18 &  -26.86 \\ 
2120+168       & -1.16 & -0.78 & 0.42 & -1.20 &  1.45 &    0.00 &   32.94 &   35.41 &  -26.79 \\ 
2121+0522      &  0.72 &  0.33 & 0.09 &  0.32 &  2.90 &   \nodata &   34.65 &   34.44 &  -27.42 \\ 
2126-158       &  0.56 &  0.48 & 0.11 &  0.52 &  2.10 &   \nodata &   34.44 &   34.48 &  -28.76 \\ 
2136+141       & -0.06 &  0.35 & 0.09 &  0.35 &  3.17 &   54.05 &   34.74 &   34.96 &  -26.91 \\ 
2146-133       & -0.80 & \nodata & \nodata & -1.01 & \nodata &   \nodata &   \nodata &   35.28 &  -25.25 \\ 
2149+212       & -0.77 & -1.28 & 0.09 & -0.94 &  4.09 &    0.81 &   35.01 &   34.86 &  -25.40 \\ 
2150+053       & -0.89 & -0.52 & 0.12 & -1.11 &  1.87 &    0.01 &   33.53 &   35.26 &  -27.19 \\ 
2156+297       & -0.87 & -1.16 & 0.09 & -0.92 &  3.46 &    0.29 &   35.10 &   35.12 &  -27.19 \\ 
2158+101       & -0.55 & -1.34 & 0.11 & -0.98 &  3.37 &   \nodata &   34.92 &   34.50 &  -26.96 \\ 
2222+051       & -0.93 & \nodata & \nodata & -1.08 &  3.27 &    0.35 &   34.79 &   35.25 &  -26.80 \\ 
2223+210       & -0.47 & -0.49 & 0.09 & -0.49 &  3.74 &   46.36 &   35.21 &   35.20 &  -26.74 \\ 
2248+192       & -1.00 & \nodata & \nodata & -0.99 & \nodata &   \nodata &   \nodata &   35.05 &  -26.25 \\ 
2249+185       & -0.79 & -0.12 & 0.09 &  0.01 &  3.08 &   \nodata &   34.34 &   35.33 &  -26.23 \\ 
2251+244       & -0.70 & -0.56 & 0.09 & -0.52 &  3.58 &    1.00 &   35.38 &   35.52 &  -27.52 \\ 
2254+0257      &  0.61 &  0.51 & 0.09 &  0.52 &  1.76 &   \nodata &   33.75 &   33.84 &  -28.01 \\ 
2338+042       & -0.97 & \nodata & \nodata & -1.07 &  3.96 &    0.50 &   35.19 &   35.63 &  -26.06 \\ 
2345+061       & -0.52 & -1.09 & 0.09 & -0.72 &  3.24 &    1.10 &   34.76 &   34.58 &  -26.90 \\ 
2351-154       &  0.09 & -0.09 & 0.09 & -0.10 &  2.70 &   \nodata &   34.96 &   34.82 &  -28.63 \\ 
2354+144       & -0.70 & -0.86 & 0.10 & -1.23 &  2.62 &    0.01 &   34.06 &   35.02 &  -26.67 \\ 
\enddata
\end{deluxetable}

\begin{deluxetable}{lrrrrrrr}
\tablewidth{0pt}
\tablecaption{Absorption Line Properties \label{tab:tab3}}
\tablehead{
\colhead{} &
\multicolumn{3}{c}{$\pm5000\,{\rm km\,s^{-1}}$} &
\multicolumn{4}{c}{$5000\,{\rm km\,s^{-1}}$ -- $55,000\,{\rm km\,s^{-1}}$} \\
\colhead{Sample} & 
\colhead{$dN/d\beta$} &
\colhead{$\sigma$} & 
\colhead{N} &
\colhead{$dN/d\beta$} &
\colhead{$\sigma$} & 
\colhead{N (obs)} &
\colhead{N (exp)}
}
\startdata
All QSOs & 12.30 & 1.06 & 126 & 5.60 & 0.37 & 224 & \nodata \\
Sample A Steep & 14.06 & 2.12 & 39 & 3.03 & 0.55 & 30 & 55 \\
Sample A Flat & 8.41 & 1.91 & 19 & 6.95 & 0.80 & 76 & 61 \\
Sample B Steep & 15.12 & 3.51 & 17 & 2.74 & 0.76 & 13 & 22 \\
Sample B Flat & 12.86 & 3.10 & 17 & 7.78 & 1.09 & 51 & 36 \\
Sample C Steep & 15.31 & 3.49 & 18 & 4.45 & 0.91 & 23 & 29 \\
Sample C Flat & 11.46 & 3.04 & 14 & 6.97 & 1.10 & 40 & 32 \\
Sample D Steep+Steep & 14.38 & 3.93 & 12 & 2.57 & 0.86 & 9 & 17 \\
Sample D Flat+Flat & 9.87 & 3.28 & 9 & 8.13 & 1.35 & 36 & 25 \\
Sample E Lo-$R_V$ & 11.87 & 2.51 & 21 & 4.67 & 0.77 & 37 & 44 \\
Sample E Hi-$R_V$ & 15.74 & 3.09 & 22 & 6.49 & 1.01 & 41 & 35 \\
Sample F $C_{20} < 1$ & 12.49 & 2.86 & 18 & 4.10 & 0.84 & 24 & 32 \\ 
Sample F $C_{20} > 1$ & 18.79 & 5.15 & 13 & 5.88 & 1.35 & 19 & 18 \\
\enddata
\end{deluxetable}

\end{document}